\documentclass[fleqn,usenatbib]{mnras}

\usepackage{newtxtext,newtxmath}

\usepackage[T1]{fontenc}

\DeclareRobustCommand{\VAN}[3]{#2}
\let\VANthebibliography\thebibliography
\def\thebibliography{\DeclareRobustCommand{\VAN}[3]{##3}\VANthebibliography}

\usepackage{graphicx}	
\usepackage{amsmath}	
\usepackage{longtable}

\title[Study of pulsar flux density and its variability with Parkes data archive]{Study of pulsar flux density and its variability with Parkes data archive}

\author[Ziyang Wang et al.]{
Ziyang Wang,$^{1,2}$
Jingbo Wang,$^{3,4}$\thanks{E-mail: 1983wangjingbo@163.com}
Na Wang,$^{2,4}$\thanks{E-mail: na.wang@xao.ac.cn}
Shi Dai$^{5,6}$
and Jintao Xie$^{2,4}$
\\
$^{1}$College of Physics Science and Technology, Xinjiang University, Xinjiang 830046, People's Republic of China\\
$^{2}$Xinjiang Astronomical Observatory, 150, Science-1 Street, Urumqi 830011, People's Republic of China\\
$^{3}$Institute of Optoelectronic Technology, Lishui University, Lishui, Zhen323000, China\\
$^{4}$Xinjiang Key Laboratory of Radio Astrophysics, 150 Science 1-Street, Urumqi, Xinjiang, 830011, People's Republic of China\\
$^{5}$Western Sydney University, Locked Bag 1797, Penrith South DC, NSW 1797, Australia\\
$^{6}$Australia Telescope National Facility, CSIRO, Space and Astronomy, PO Box 76, Epping, NSW 1710, Australia
}

\date{Accepted 2023 January 16. Received 2023 January 15; in original form 2022 May 09}

\pubyear{XXX}

\begin{document}
\label{firstpage}
\pagerange{\pageref{firstpage}--\pageref{lastpage}}
\maketitle

\begin{abstract}
We present average flux density measurements of 151 radio pulsars at 1.4 GHz with the Parkes `Murriyang' radio telescope.
We recommend our results be included in the next version of the ATNF pulsar catalogue. 
The large sample of pulsars together with their wide dispersion measure (DM) range make this data set useful for studying variability of flux density, pulsar spectra, and interstellar medium (ISM). 
We derive the  modulation indices and  structure-function from the flux density time series for 95 and 54 pulsars, respectively. We suggest the modulation index also be included in the next version of the pulsar catalogue to manifest the variability of pulsar flux density. 
The modulation index of flow density and DM are negatively correlated.
The refractive scintillation (RISS) timescales or its lower bound for a set of 15 pulsars are derived. 
They are very different from theoretical expectations, implying the complicated properties of the ISM along  different lines of sight.
The structure-function for other pulsars is flat. The RISS parameters for some of these pulsars possibly could be derived with different observing strategies in the future.
\end{abstract}



\section{INTRODUCTION} 
\label{INTRODUCTION}

More than 50 years after discovering pulsars \citep{1968Natur.217..709H}, it is still not possible to give a definitive exposition of the processes by which pulsars emit beams of radio waves. 
Generally, the pulsar radio emission is described using models that include a magnetosphere filled with an electron-positron plasma that corotates with the pulsar.
However, important details such as the location of the emission sites and the emission mechanism are still confusing. Studying flux density on one or more frequencies can assist in developing an understanding of the pulsar emission mechanism \citep[e.g.][]{2018MNRAS.473.4436J,2016RAA....16..159H,2019ApJ...874...64Z}. {Meanwhile, measurement of flux density can provide a method to probe the interstellar medium \citep[e.g.][]{2018MNRAS.474.4637K}.}

However, accurate data on flux density are lacking for the majority of pulsars.  The ATNF Pulsar Catalogue \citep[version 1.67;][]{2005AJ....129.1993M} shows that pulsar flux densities are relatively well known nearby 1.4 GHz and 400 MHz, where a majority of the pulsars were discovered but are not well known at other frequencies. 
Out of 3319 known pulsars, about 69$\%$ have flux density measurements at 1.4 GHz; above 2 GHz, the fraction is only 18$\%$; and between 600 to 900 MHz, there are flux density measurements for only 33$\%$ of the pulsars. Furthermore, about 61$\%$ of flux density measurements at 1.4 GHz and nearly 80$\%$ of recorded flux density measurements at 800 MHz are obtained from the first discovery articles, which are usually not absolute calibrated and only estimated with radiometer equation and parameters known to the observing system \citep[e.g.][]{2006MNRAS.372..777L}. This way greatly saves the observing time of the noise diode and calibrator source before each observation, but there are significant differences in the flux density obtained from multiple measurements \citep[e.g.][]{2013MNRAS.434.1387L}. The information of accurate flux density measurements can be used to make accurate predictions for pulsar surveys and observations with the Square Kilometre Array \citep[SKA;][]{2015aska.confE..40K}, FAST, and other radio telescopes. The flux density measurements could also contribute to optimising observing strategies or to the design of surveys using pulsar population synthesis \citep[e.g.][]{2014MNRAS.439.2893B}. 

The flux density of pulsars shows variability on a variety of time scales. The variability is because of a combination of extrinsic due to the propagation of the radio  emission through the ISM and intrinsic to the emission mechanism. The ISM is inhomogeneous on a wide range of length scales. Inhomogeneities in the ISM cause scattering of the radio waves. This is responsible for a variety of related phenomena:  temporal smearing of pulsed signals, apparent angular broadening of pulsar images, and diffractive interstellar scintillation \citep[DISS; e.g.][]{2005MNRAS.358..270W}.
DISS displays a short time scale variation of pulsar flux density caused by the inhomogeneity of electron density in the ISM. The interference pattern changes with time due to the relative motion between the pulsar, scattering screen, and observer. It leads to drastic variations in observed flux density with a timescale from several minutes to several hours \citep{1995ApJ...443..209A}. The long timescale variation of flux density from days to months is the result of refractive scintillation due to the refraction scattering of electron density in the interstellar medium at large spatial scales \citep{1984A&A...134..390R}. Pulsars with high dispersion measure at large distances have stable observed flux densities over the years \citep{2000ApJ...539..300S}, which indicate the emission from pulsars is stable when single pulses are integrated hundreds of times, and diffractive scintillation has been taken into account.

Soon after discovering the pulsar, DISS was found in pulsars {\citep{1969Natur.221..158R}}, and a host of related literatures have been published {\citep[e.g.][]{2013MNRAS.429.2161K}}. \cite{1985ApJ...288..221C} studied the interstellar scintillation observation of pulsars at low frequency (near the frequency of 400 MHz) and provided the observation parameters and de-correlation bandwidth parameters of 36 pulsars, as well as the dynamic spectrum of 4 pulsars with their characteristics. The Nanshan 25 meters radio telescope obtained the dynamic spectrum of 7 pulsars at the frequency of 1540 MHz \citep{2001ChJAA...1..421W}. The  observational evidence of RISS in pulsars \citep{1982A&A...113..311S} shows that DM is correlated with long-term flux density changes. Since then, several investigations on RISS have been conducted \citep[e.g.][]{2000ApJ...539..300S,2021MNRAS.501.4490K}. {But the RISS phenomenon has been difficult to explain because how inhomogeneities of ionized ISM in the Galaxy are distributed is not well known.} {DISS and RISS phenomenon have been studied at different frequencies. However, long-term observations are essential to investigation of RISS. Most of young pulsars are regular monitored around 1.4 GHz. Pulsars are in the regime of strong scintillation except very close pulsars at 1.4 GHz.}



This work presents flux density measurements of 151 pulsars around 1.4 GHz with the Parkes `Murriyang' radio telescope. We describe the observational system, data set, and data reduction in Section~\ref{OBSERVATIONS}. Our main results are presented in Section~\ref{RESULTS}. 
Section~\ref{DISCUSSION} compares our results with theoretical expectations and discusses their implications.
Finally, we summarize our results in Section~\ref{CONCLUSIONS}.

\section{OBSERVATIONS and DATA PROCESSING}
\label{OBSERVATIONS}

The CSIRO pulsar data archive \citep{2011PASA...28..202H} includes most of the pulsar observations made with the Parkes telescope (the earliest observations in the archives began in 1991). After an 18-month embargo, the data will be publicly available. 
The central frequency of all the selected observations is close to 1400 MHz. We include data taken with the H-OH receiver \citep{1990ITAP...38.1898T} and the multibeam receiver \citep{1996PASA...13..243S}. The majority of data were recorded with the Parkes Digital Filter Bank system (PDFB) with  256 MHz bandwidth centred at 1369 MHz and 1024 frequency channels for the multibeam receiver and the H-OH receiver. Details of receivers and backends can be found in \cite{2013PASA...30...17M}. All data were recorded with the PSRFITS data format \citep{2004PASA...21..302H}. 

We investigated a total of 151 pulsars in the CSIRO pulsar data archive. These pulsars have not been well studied for long-term flux density variability. We obtained flux density and uncertainty for these pulsars. Our sample includes young pulsars, millisecond pulsars, binary pulsars, and pulsars in globular clusters. 

Data processing procedure follows \cite{2019RAA....19..103X}, using the {\sc psrchive} software package \citep{2004PASA...21..302H}, {which includes {\sc pazi}, {\sc paas}, {\sc psrflux}, and other tools.} We first deleted data affected by narrow-band and impulsive RFI and 5\% of the band edges using {\sc paz}. We used {\sc pazi} to check the pulse profiles and deleted frequency channels or sub-integrations affected by RFI. Then, We used particular Hydra A flux calibration solutions matching with the receiver and backend instruments used. {All the data sets of Hydra A were obtained from the Parkes Pulsar Timing Array project (PPTA; \cite{2013PASA...30...17M}). The observations intervals of Hydra A are 2–3 weeks. There are multiple sets of noise diode observations during the calibration procedure, and we choose the one with the closest time. The pulsar observations were calibrated with their associated calibration files using the {\sc pac} to flatten the bandpass, transform the polarization products to Stokes parameters, and calibrate the pulse profiles in flux density units of Jansky ($\rm Jy = {10}^{-26} W/({m}^{2} \cdot Hz)$).}
Observations without calibration data were identified and deleted. We used {\sc paas} to form analytical templates from observations and then used {\sc psrflux} to acquire the flux density.

\section{RESULTS} 
\label{RESULTS}

The data set of 151 pulsars represents a large sample to study pulsar flux density. It includes pulsars with a wide range of periods, DM, and distance. The data span for a certain pulsar can be as short as a few days or as long as 14 years. About half of these pulsars were observed for more than two years.  About one-third of pulsars  were observed more than 30 times. Twenty-three pulsars were observed more than 50 times. Table~\ref{longtable} provides the pulsar name, rotation period (P), DM, distance (Dist), the span of the observations (T), and the number of observations ($N_{\rm obs}$).
\begin{figure}
	\centering
	\includegraphics[width=\columnwidth]{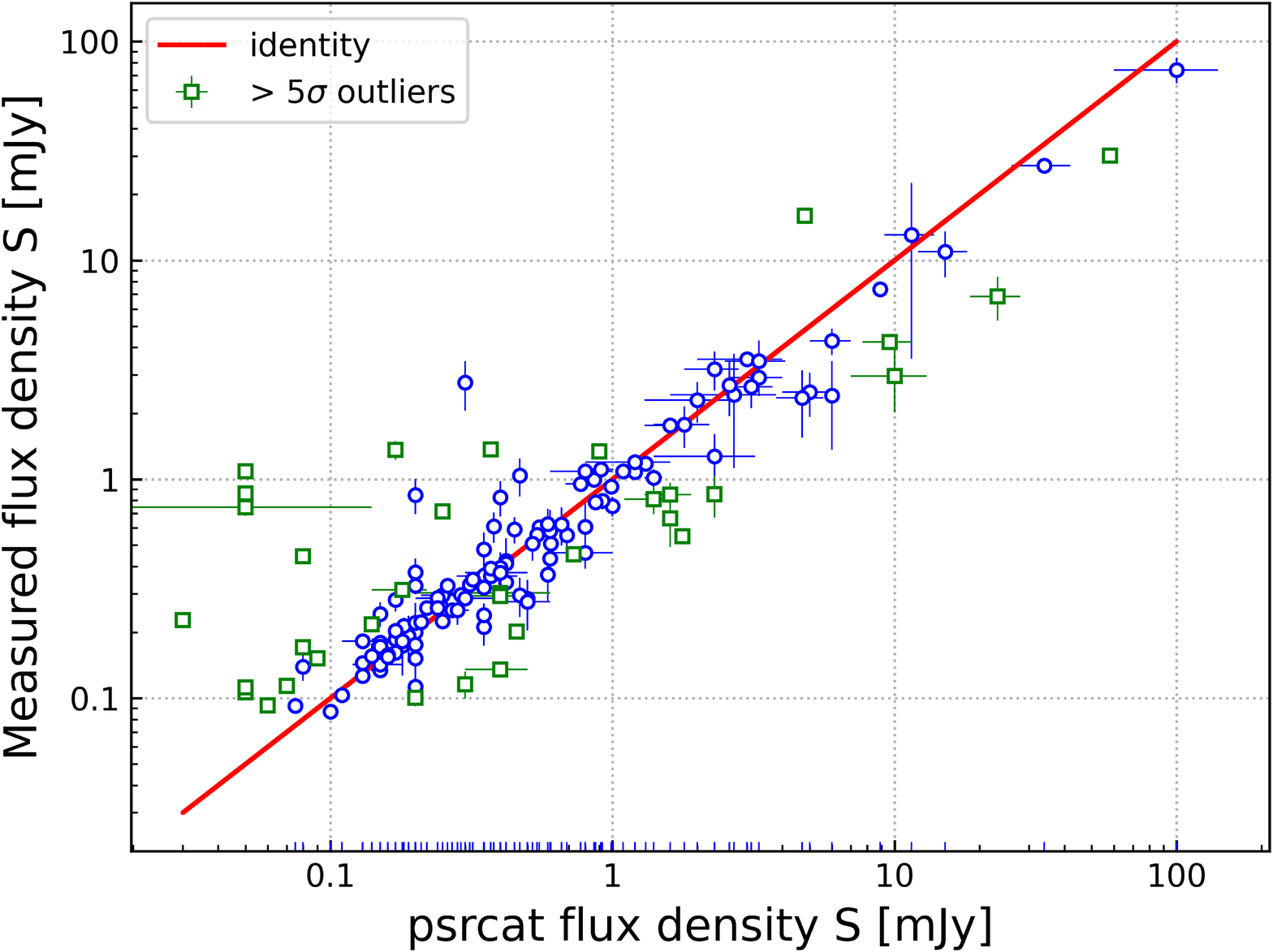}
	\caption{Comparison between the measured average flux densities of this work and flux densities within 100~MHz at a centre frequency of 1.4~GHz from the ATNF pulsar catalogue. The uncertainty of measured average flux density is estimated using Equation \ref{eq:TotalFluxDensityUncertainty}. The red line indicates the identity.}
	\label{fig:PsrcatFlux}
\end{figure}
\begin{figure*}
	\centering
    \includegraphics[width=7cm]{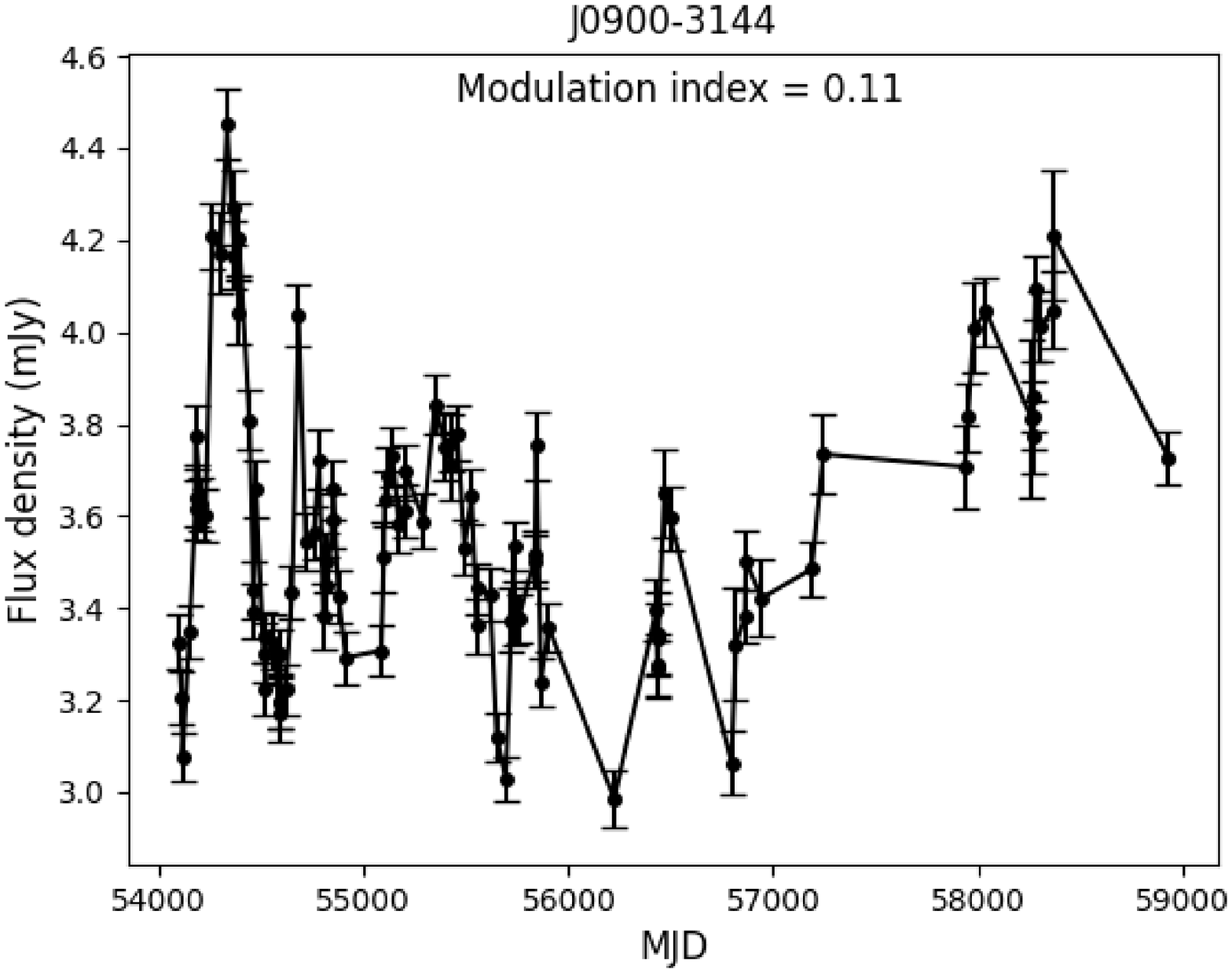}
	\includegraphics[width=7cm]{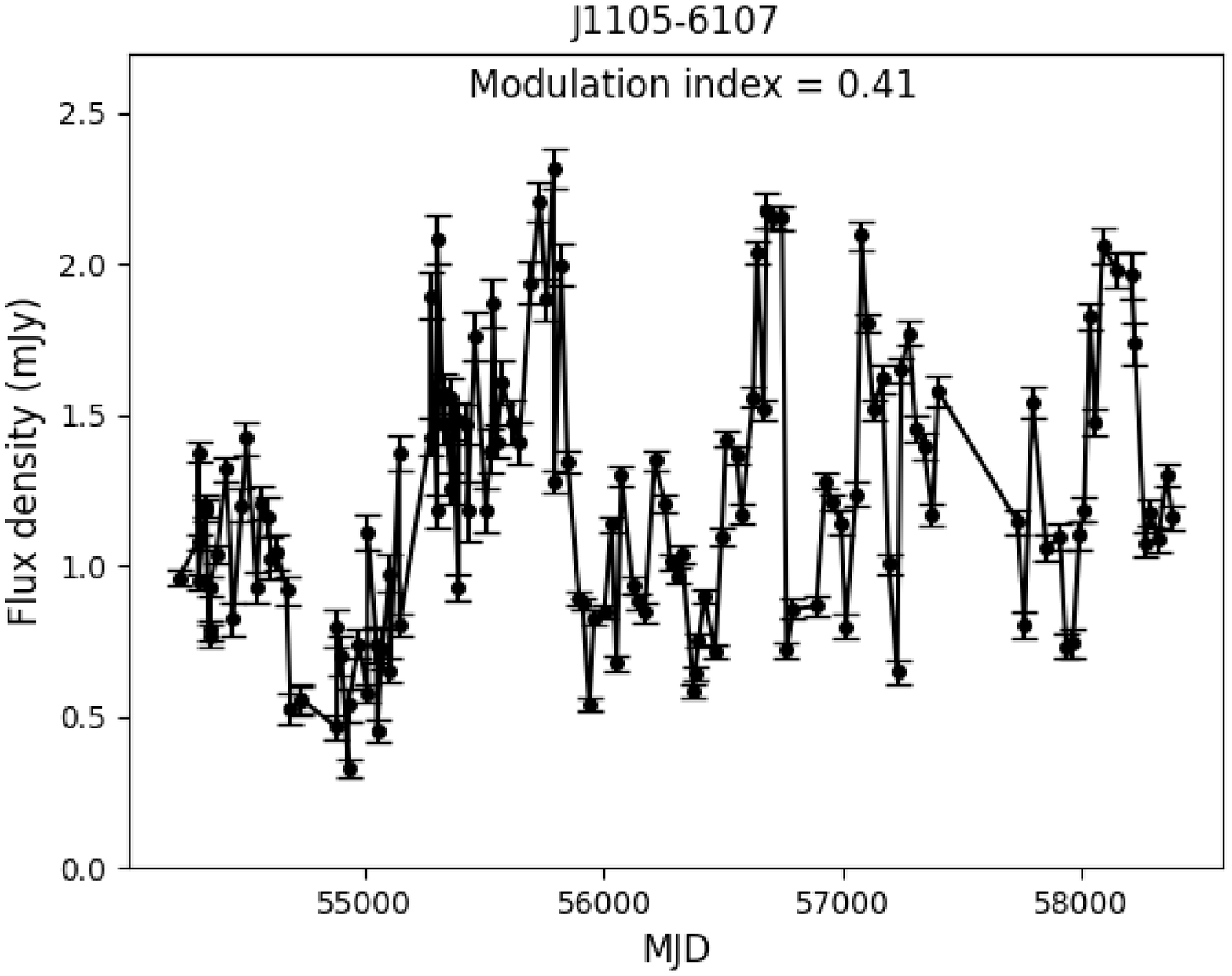}
    \includegraphics[width=7cm]{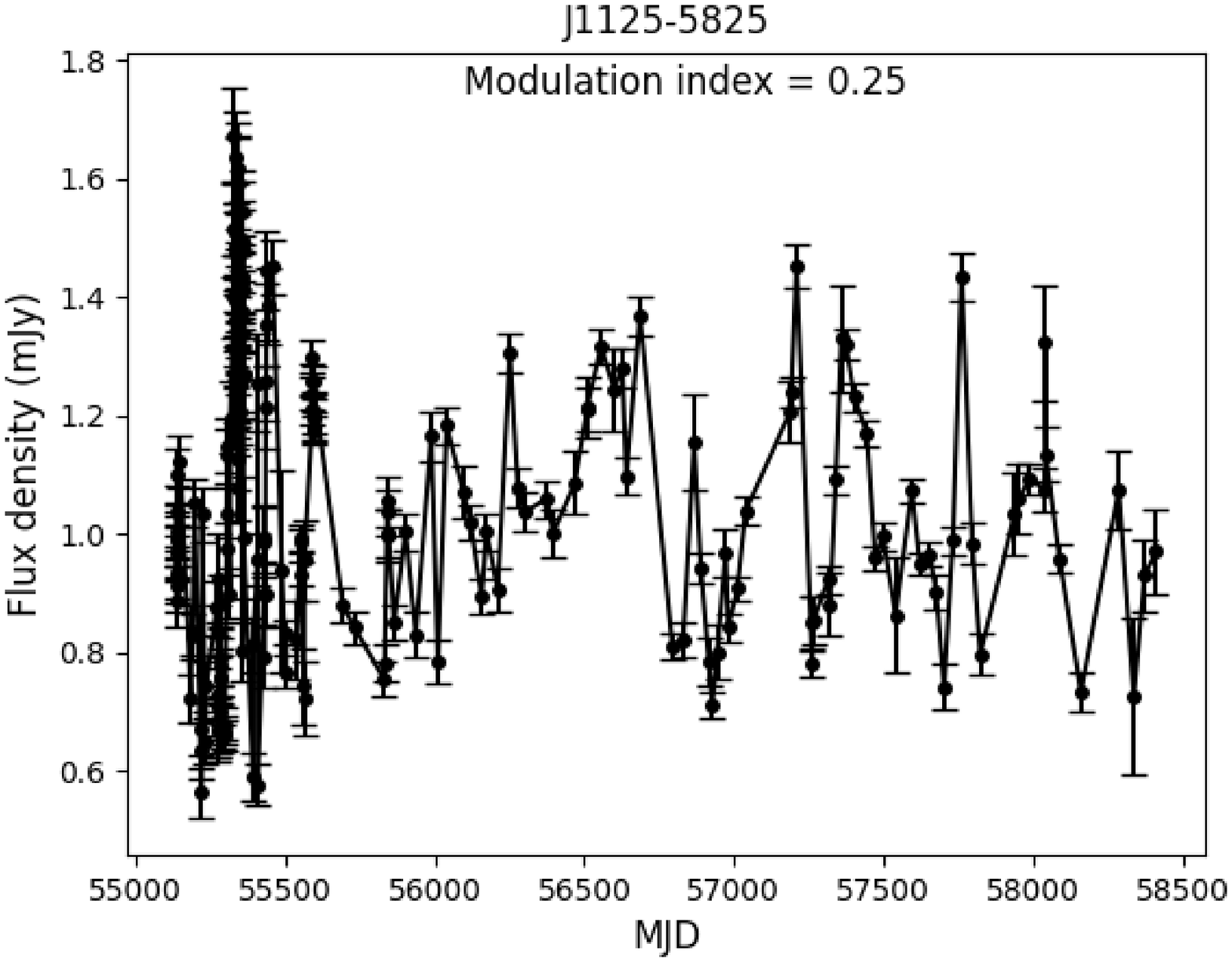}
    \includegraphics[width=7cm]{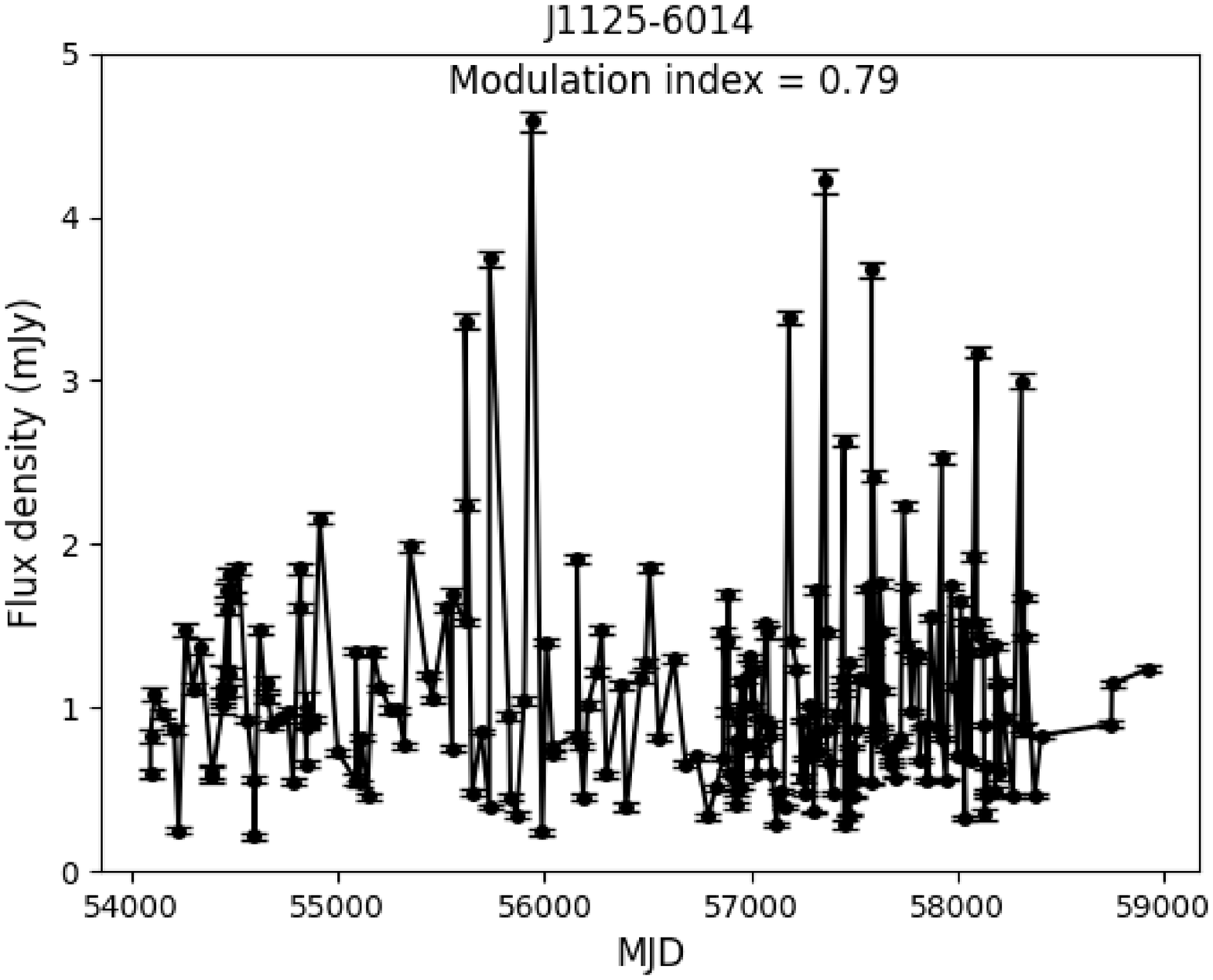}
    \includegraphics[width=7cm]{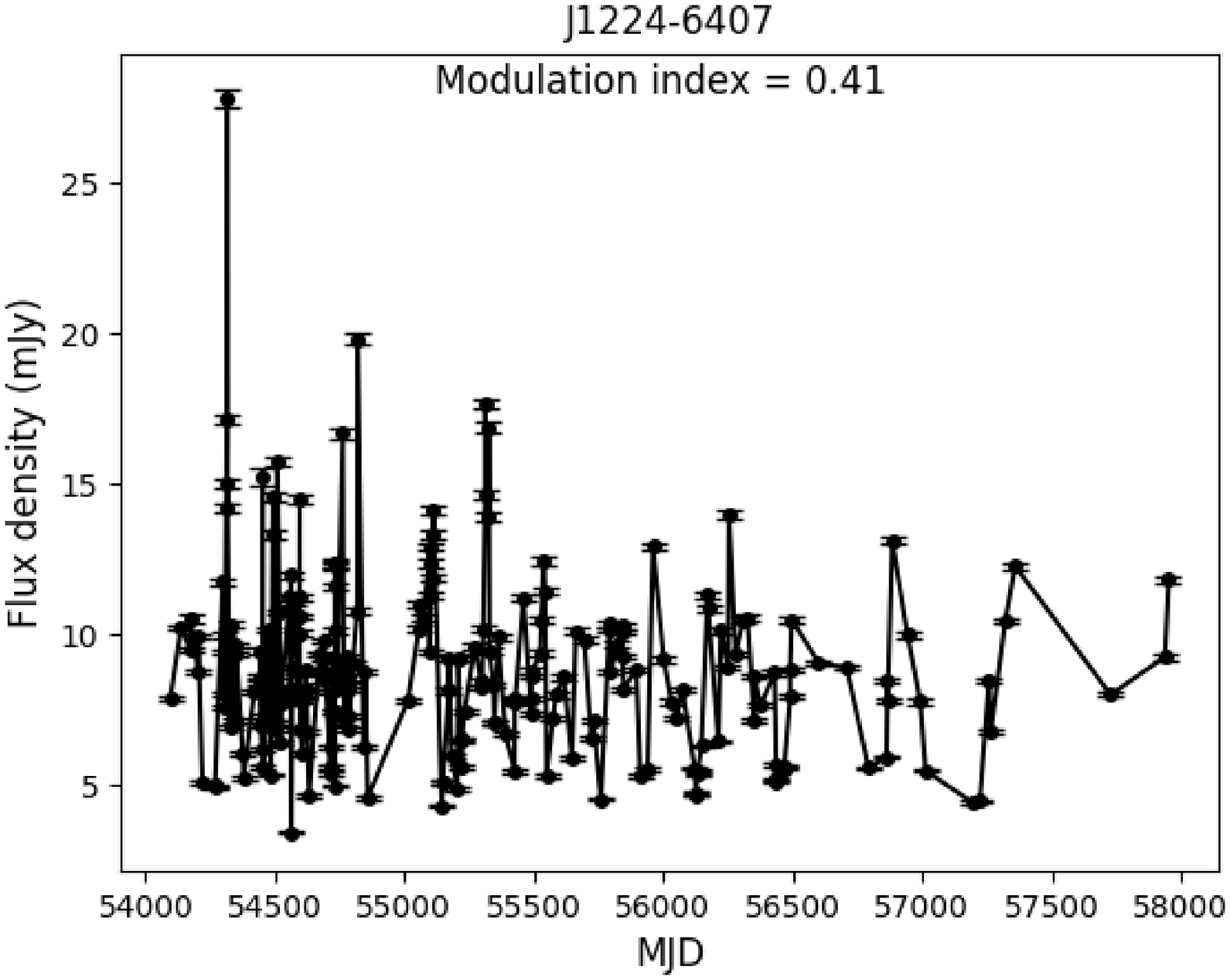}
    \includegraphics[width=7cm]{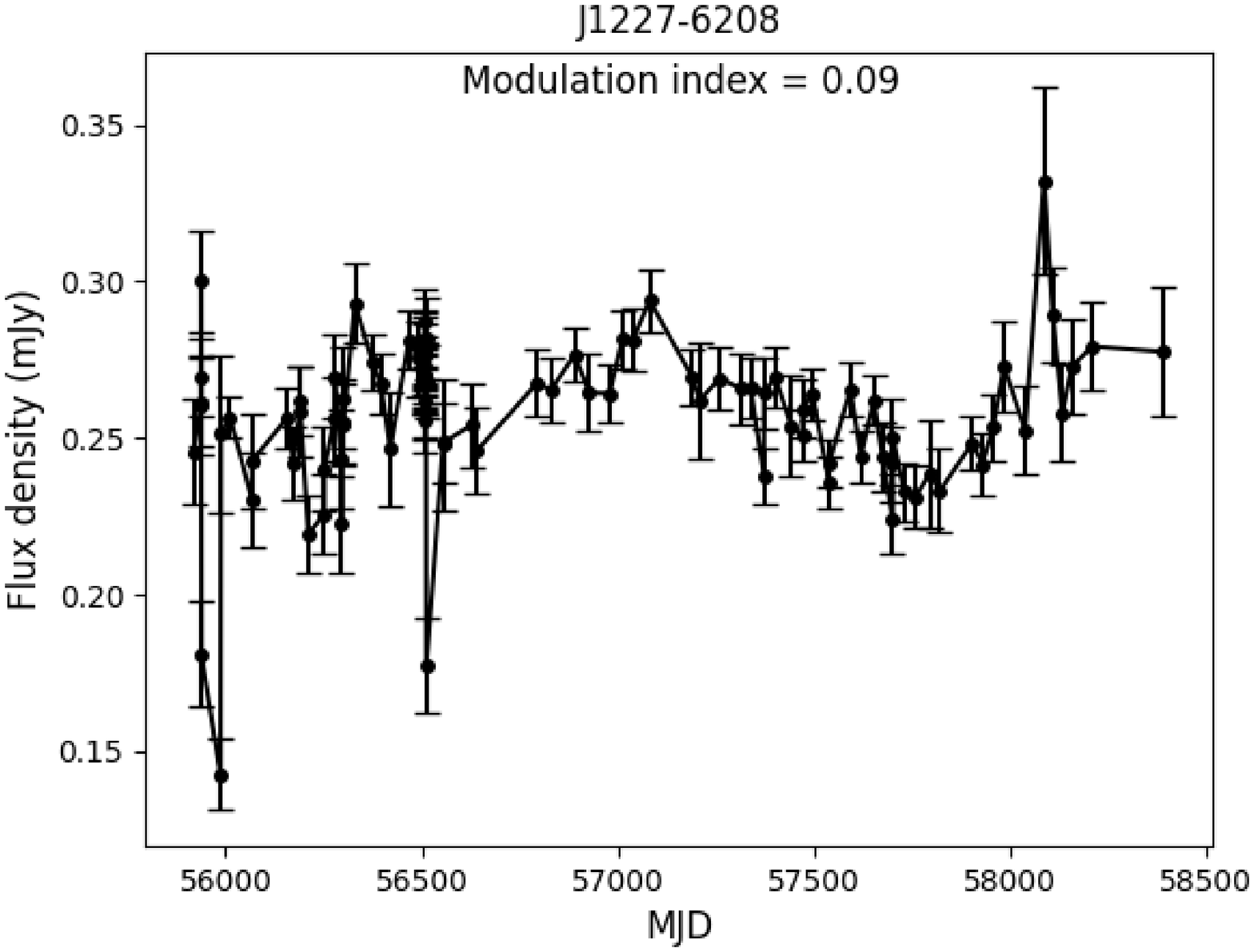}
    \includegraphics[width=7cm]{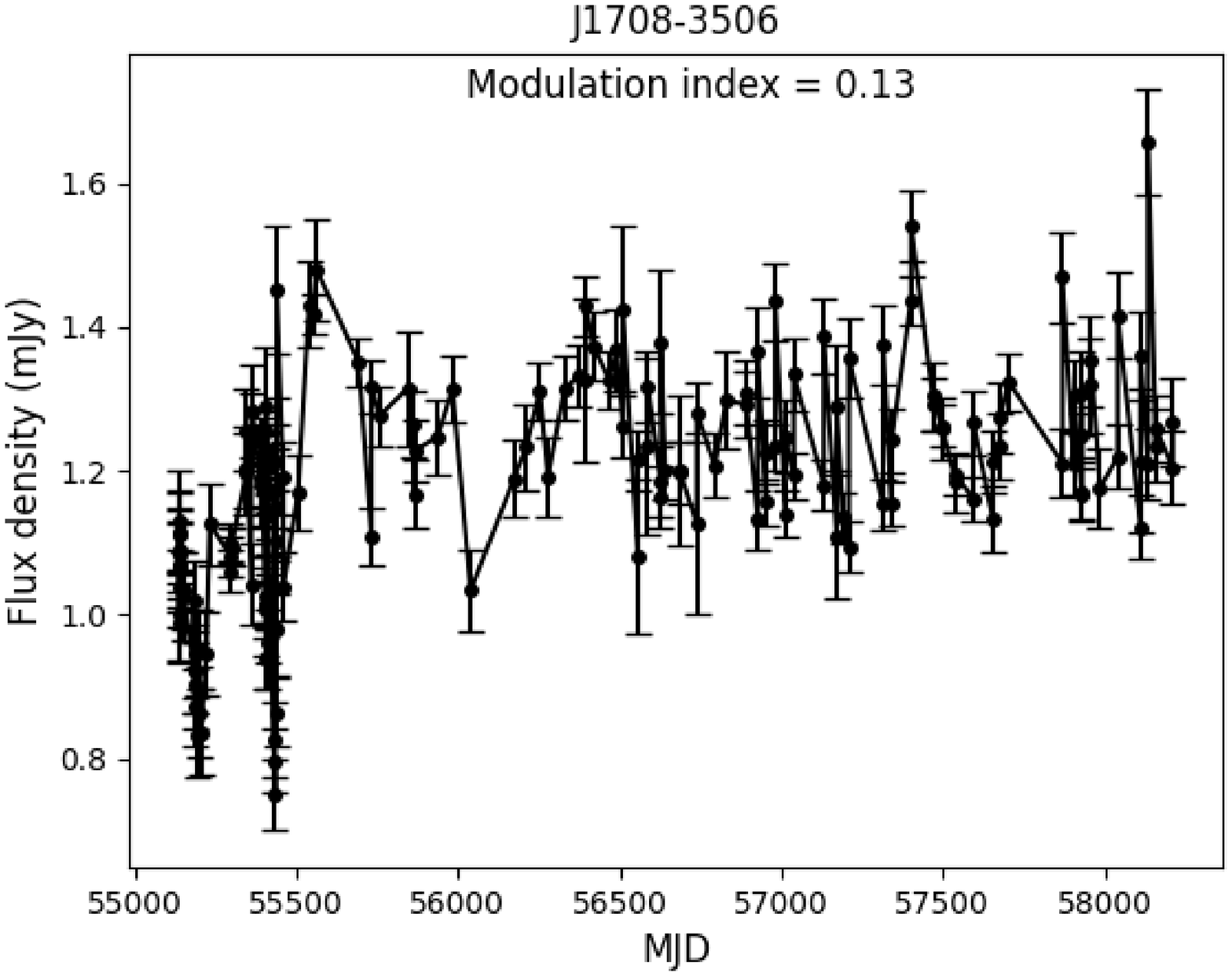}
    \includegraphics[width=7cm]{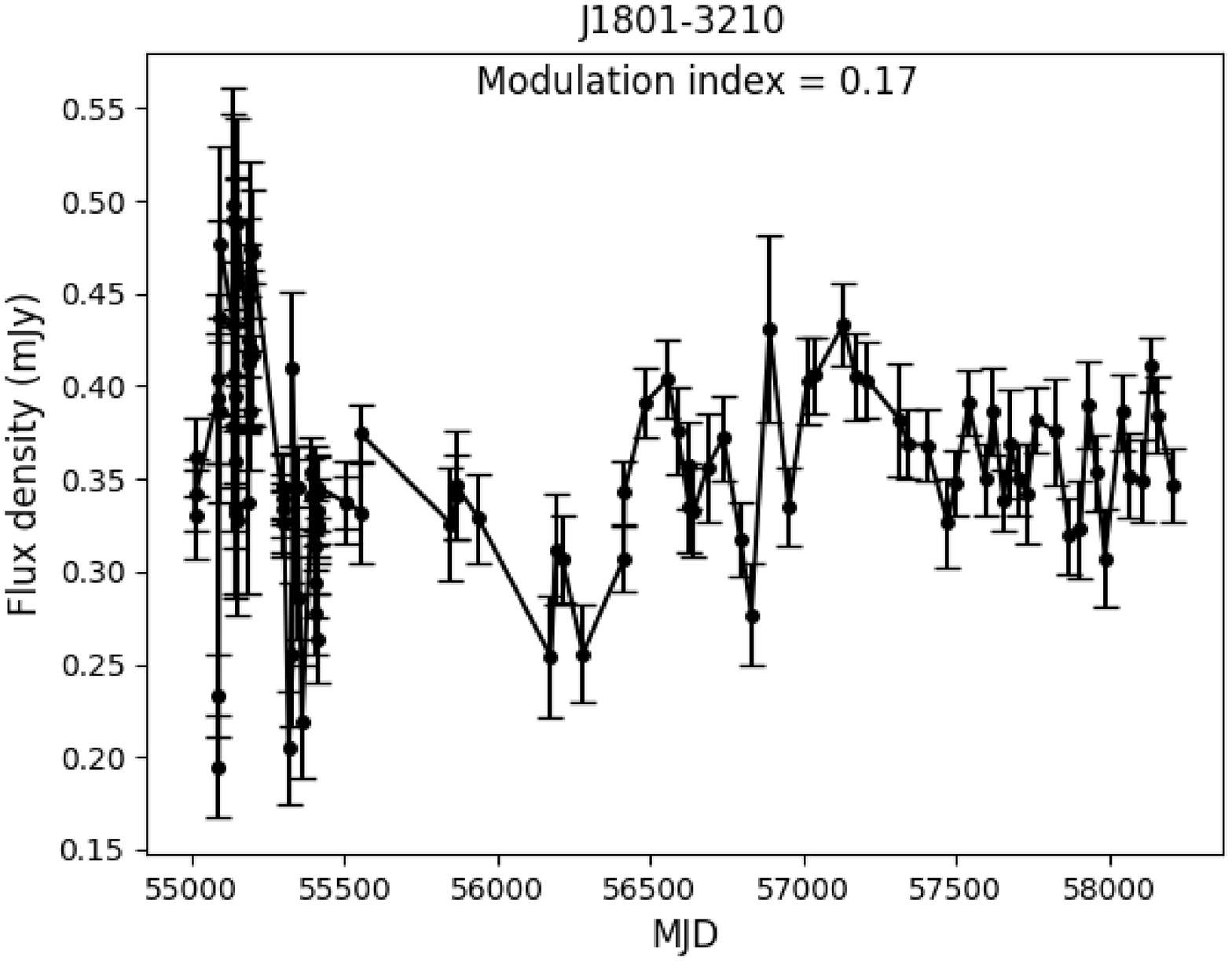}
	\caption{Flux density as a function of MJD of PSRs J0900-3144, J1105-6107, J1125-5825, J1125-6014, J1224-6407, J1227-6208, J1708-3506 and J1801-3210. The modulation index is calculated using $ m ~=~ \frac{\sigma_{S}}{\overline{S}}$. }
	\label{fig:MJD}
\end{figure*}

\subsection{Flux density}

{The flux density of pulsars, $S$, and its error, $e$, for each observation are measured using {\sc psrflux} as :
\begin{equation}
    S = (\sum_{i}^{n_{\rm on}} I_{\rm i} ) / n_{\rm tot}
\end{equation}}

{and 
\begin{equation}
    e = \sigma \sqrt{(n_{\rm on})} / n_{\rm tot},
\end{equation}
where $I_{\rm i}$ is the flux density of i phase bin, $\sigma$ is the root-mean-square of the `off-pulse' profile, $n_{\rm tot}$ is the total amount of phase bins in each period, and $n_{\rm on}$ is the number of phase bins in the `on' part of the profile.} The measured flux densities of pulsars may be greatly affected by interstellar scintillation. Therefore, it is crucial to consider its effect to estimate the reliable pulsar average flux densities and their uncertainties. We derive the variability from the flux density time series data. 
We individually obtained each observation's flux densities and uncertainties and calculated the error-weighted mean, $\overline{S}$.
Its uncertainty, $\sigma$, is derived as follows \citep{2018MNRAS.473.4436J}:
\begin{equation}
	\sigma^2 = \sigma_{\text{sys}}^2 +
	\begin{cases}
		\frac{\sigma_{\text{r}, \nu}^2}{N_{obs}} + \left( \frac{6}{5} \frac{1}{N_{obs}} - \frac{1}{5} \right) \: \sigma_{\text{scint}}^2 (\text{DM}, \nu)							& \text{if} \ 1 \leq N_{obs} < 6\\
		\frac{\sigma_{\text{r}, \nu}^2}{N_{obs}}	& \text{if} \ N_{obs} \geq 6\\
  \end{cases},
 \label{eq:TotalFluxDensityUncertainty}
\end{equation}

{where $\sigma_{\text{sys}}$ is the combined systematic uncertainty arises from the limited accuracy of the absolute flux density calibration, variations of system temperature, and other unknown factors, whose relative value is assumed to be 5\%. The uncertainty, $\sigma$, consists of the standard error obtained from the robust standard deviation $\sigma_{\text{r}, \nu}$ calculated from all measurements. The robust standard deviation $\sigma_{\text{r}, \nu}$ is computed using the interquartile range (IQR): $\sigma_{\text{r}, \nu} = 0.9183 \: \text{IQR} =  0.9183 \: \left (q_{75} (S_{\nu}) - q_{25} (S_{\nu}) \right)$. This method of analysis allows a few damaged data points without affecting the overall result.}
For pulsars with less than six observation epochs, we add to that in quadrature the uncertainty due to scintillation, $\sigma_{\text{scint}}$ is obtained using {$u_{\text{scint}} = m_{\text{r},\nu} \left( \text{DM}, \nu \right) \: \bar{S}_\nu$} \citep{2018MNRAS.473.4436J}. Columns~7 of Table~\ref{longtable} list the average flux density and its uncertainty of all the 151 pulsars. The {uncertainty} of average flux density is estimated using Equation~\ref{eq:TotalFluxDensityUncertainty}. We recommend that these average flux densities and their uncertainty be included in the next version of the pulsar catalogue.

\begin{figure*}
	\centering
	\includegraphics[width=5.3cm,height=4.1cm]{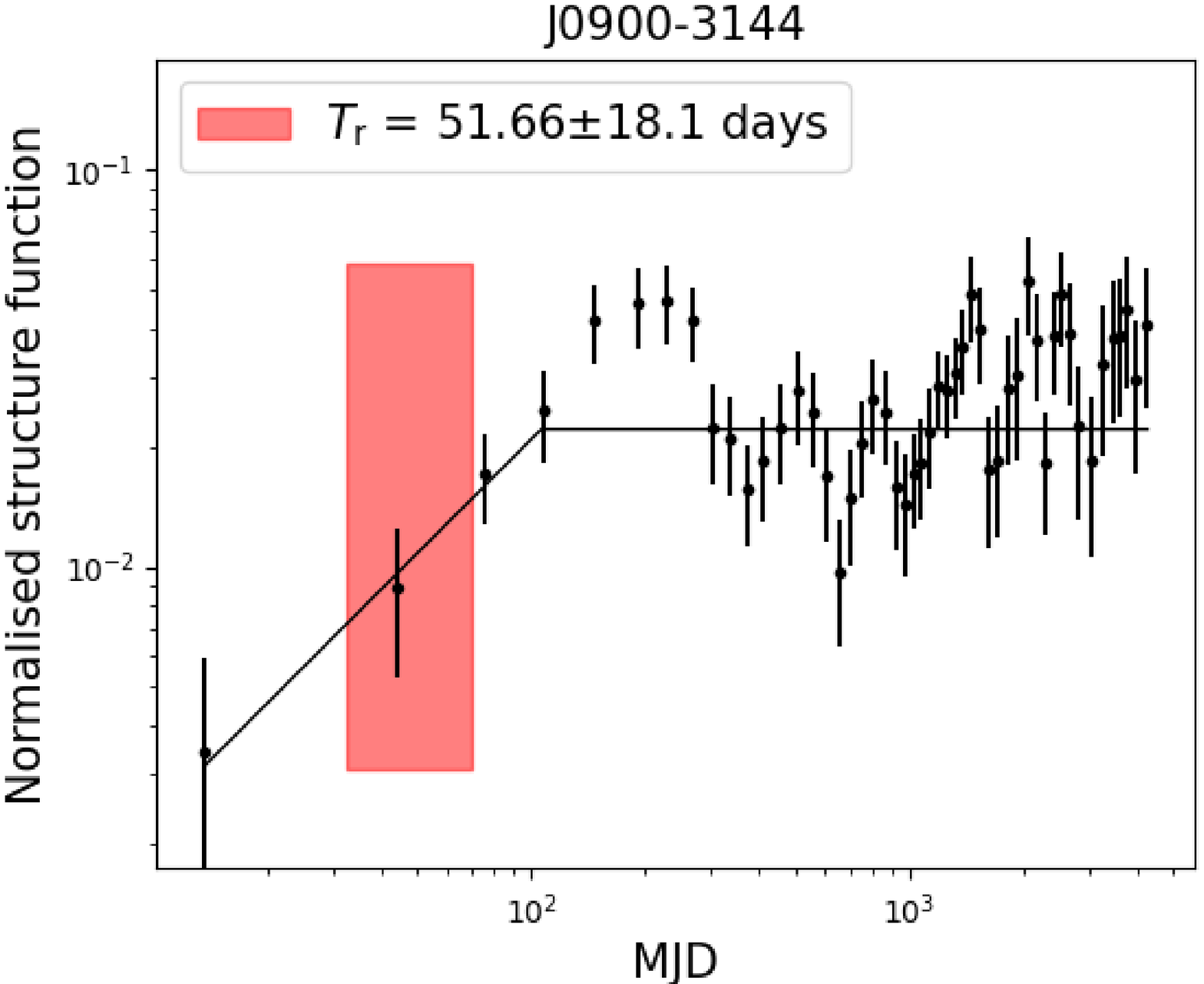}
	\includegraphics[width=5.3cm,height=4.1cm]{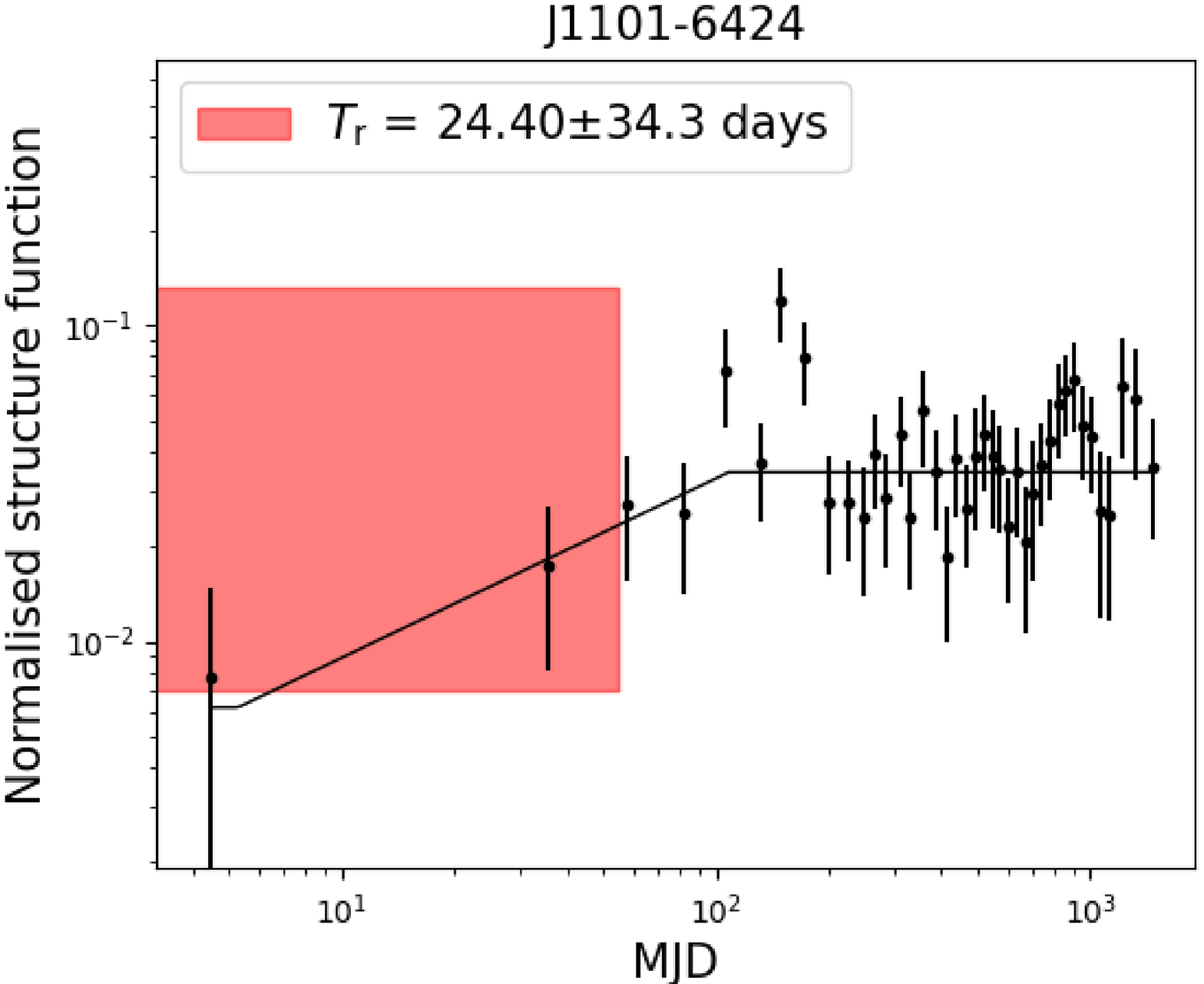}
	\includegraphics[width=5.3cm,height=4.1cm]{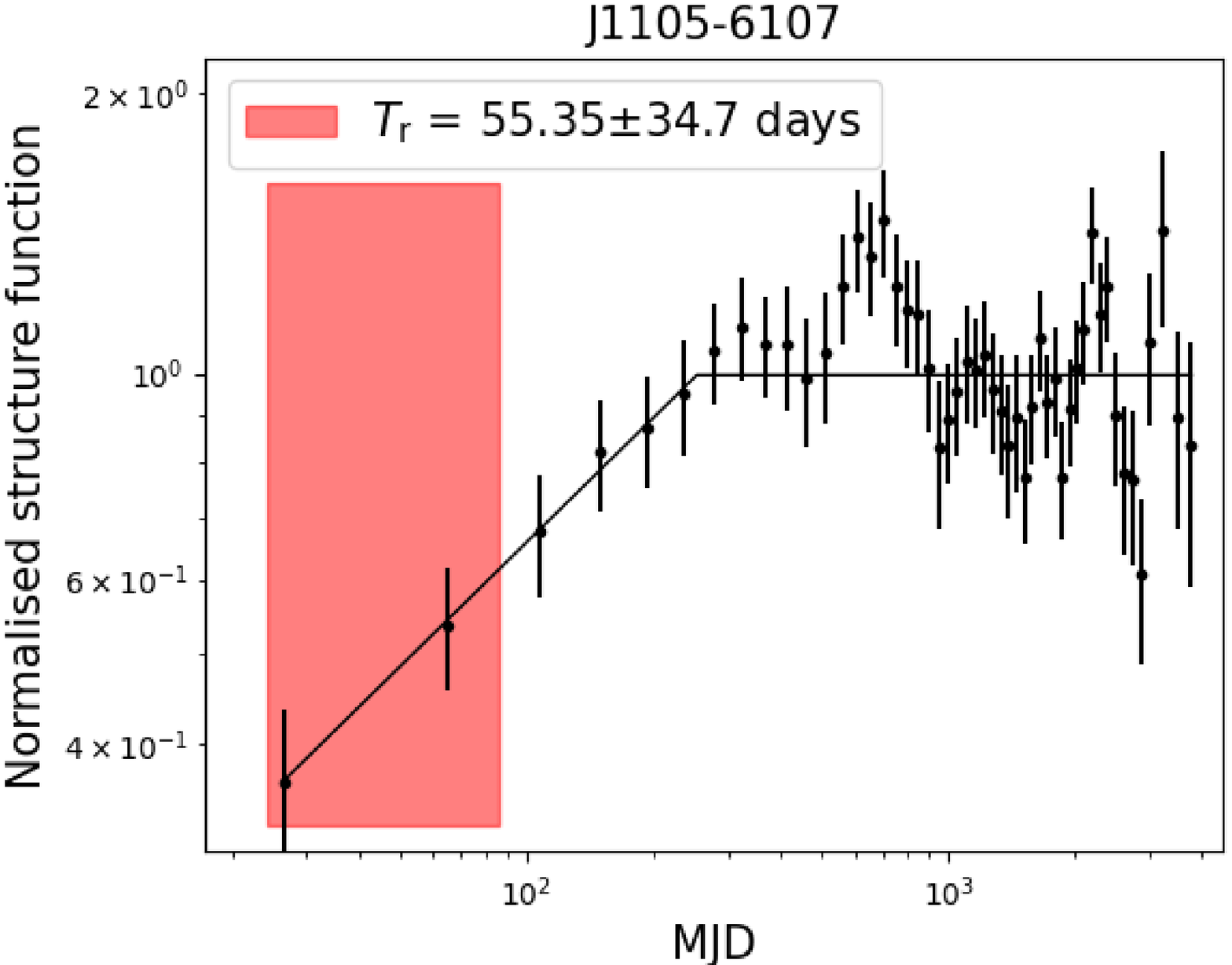}
	\includegraphics[width=5.3cm,height=4.1cm]{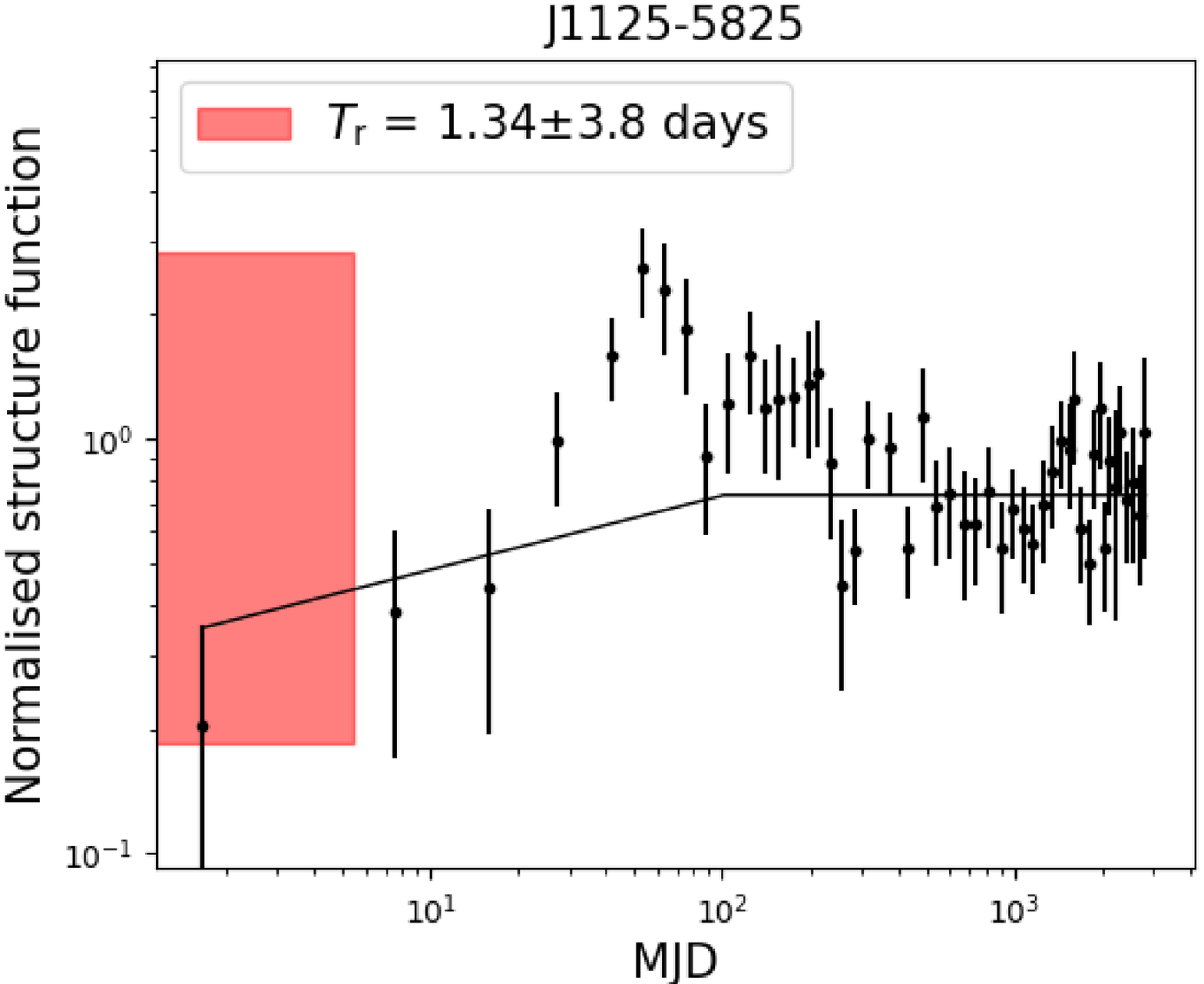}
	\includegraphics[width=5.3cm,height=4.1cm]{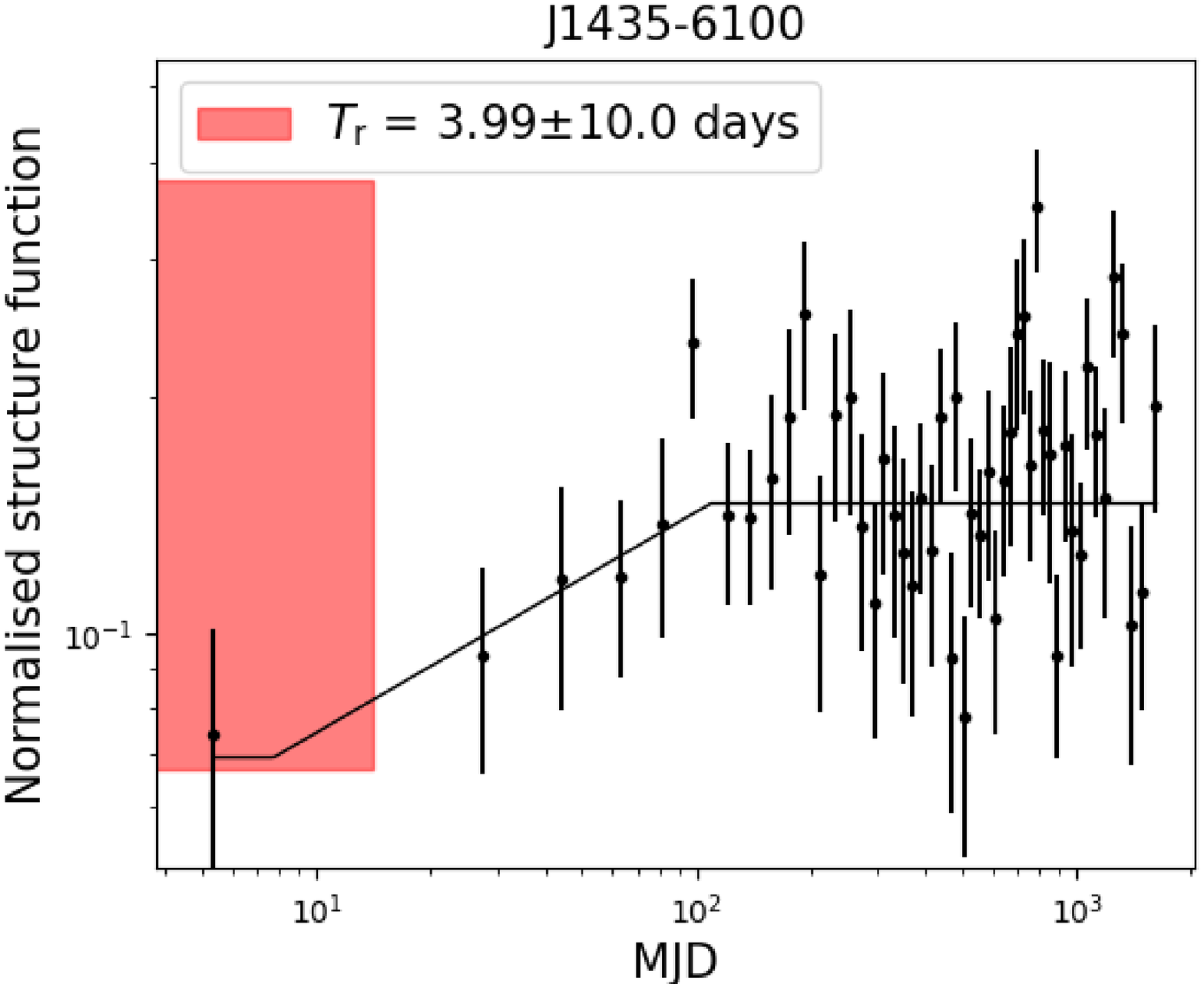}
	\includegraphics[width=5.3cm,height=4.1cm]{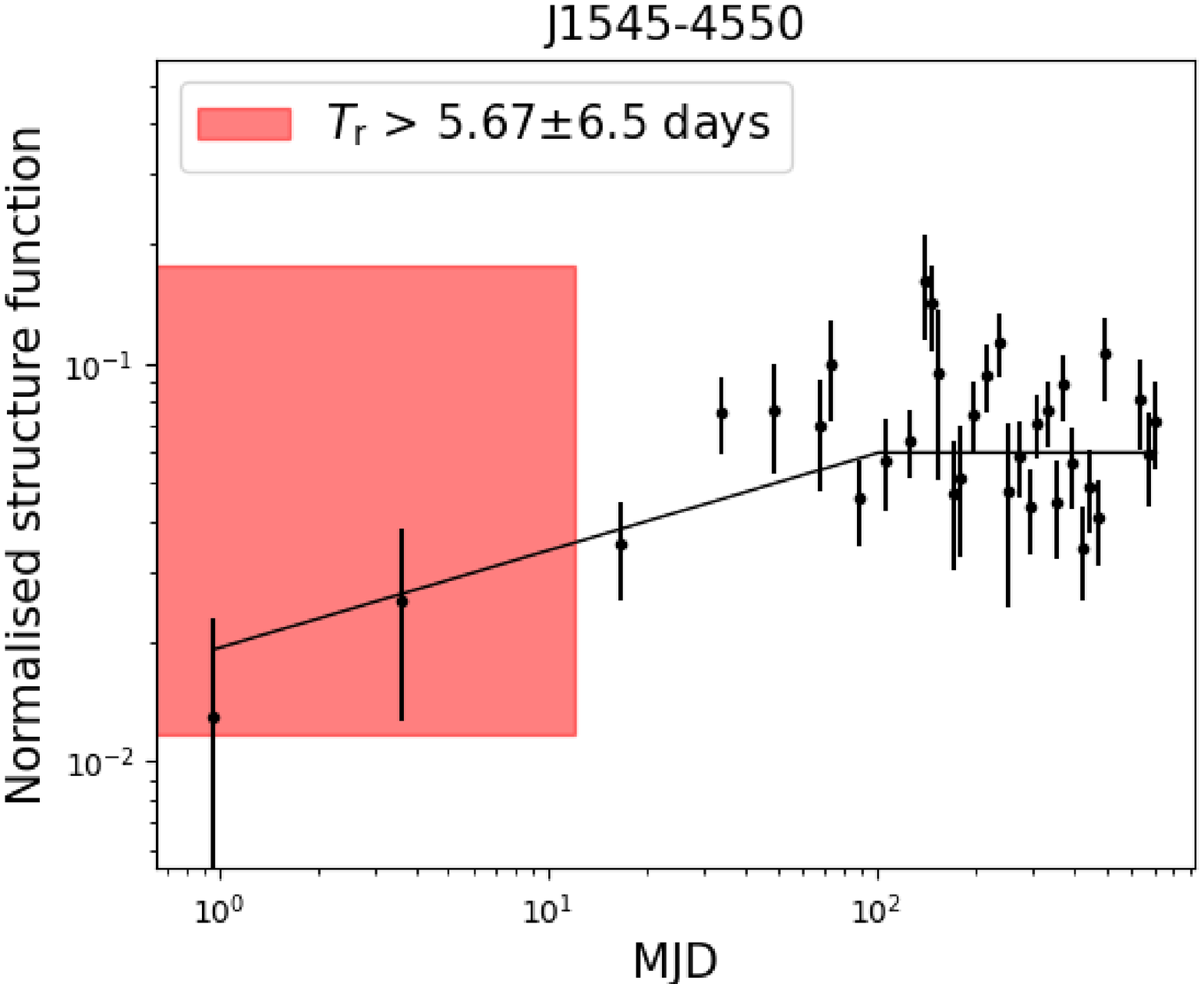}
	\includegraphics[width=5.3cm,height=4.1cm]{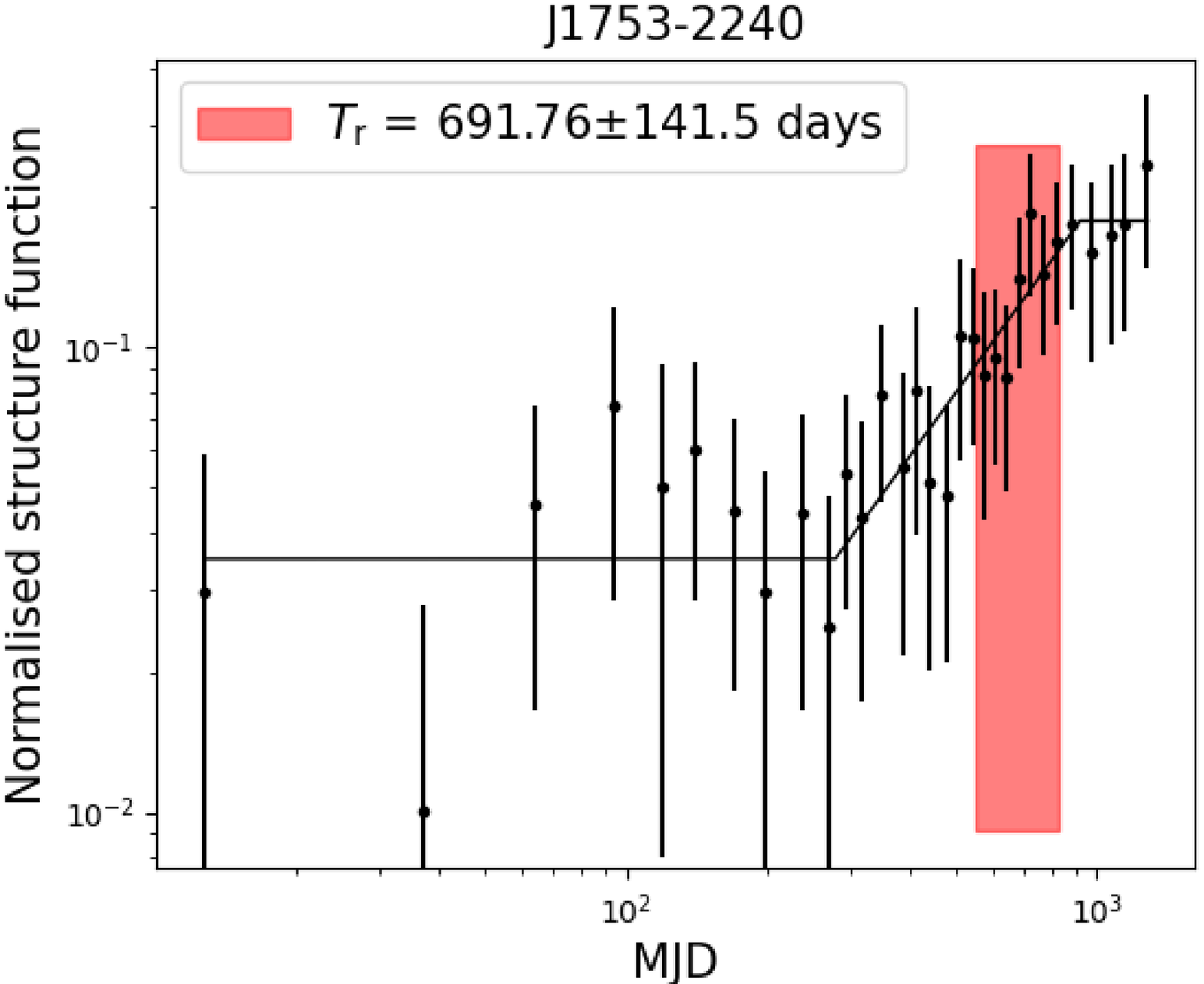}
	\includegraphics[width=5.3cm,height=4.1cm]{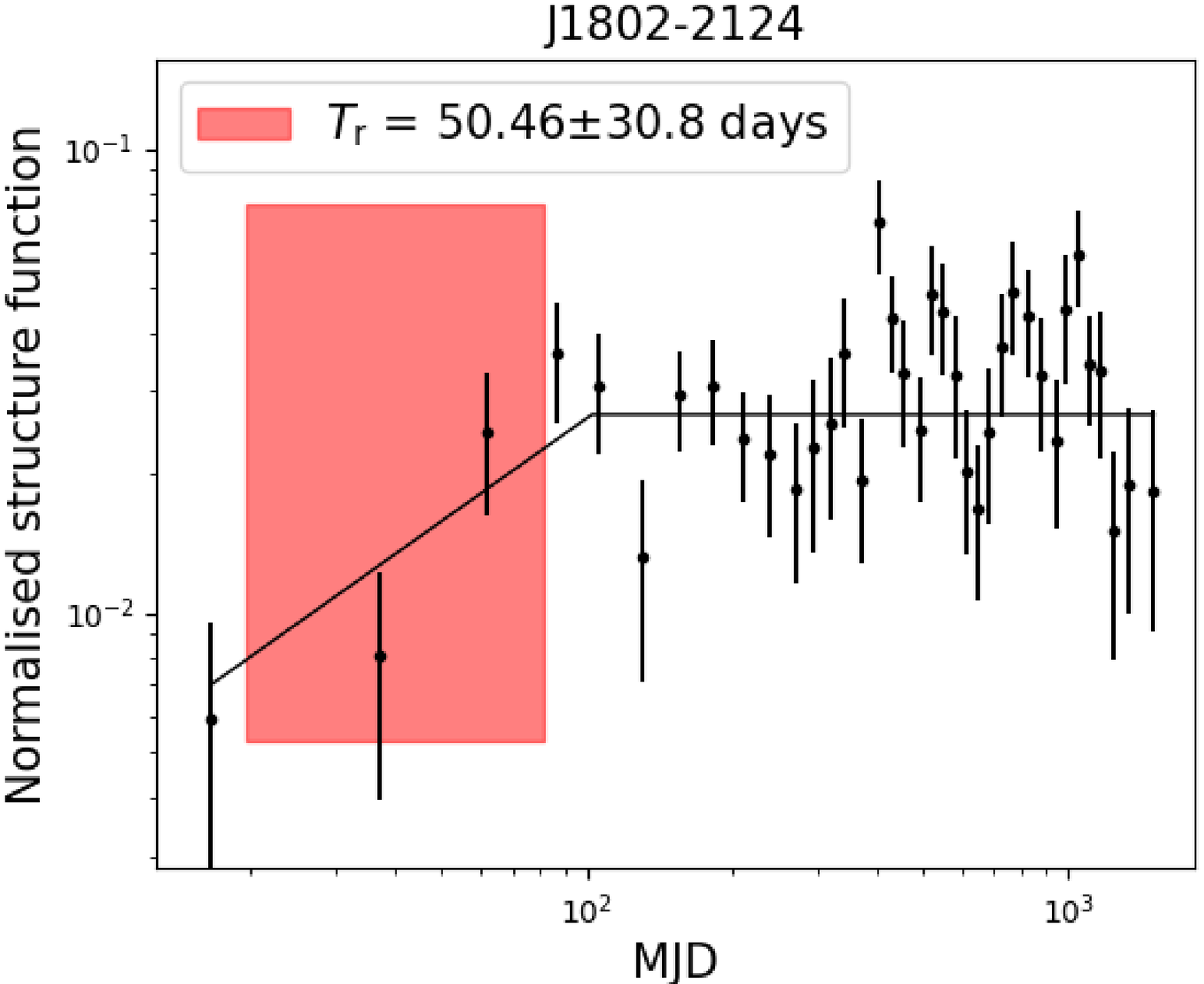}
	\includegraphics[width=5.3cm,height=4.1cm]{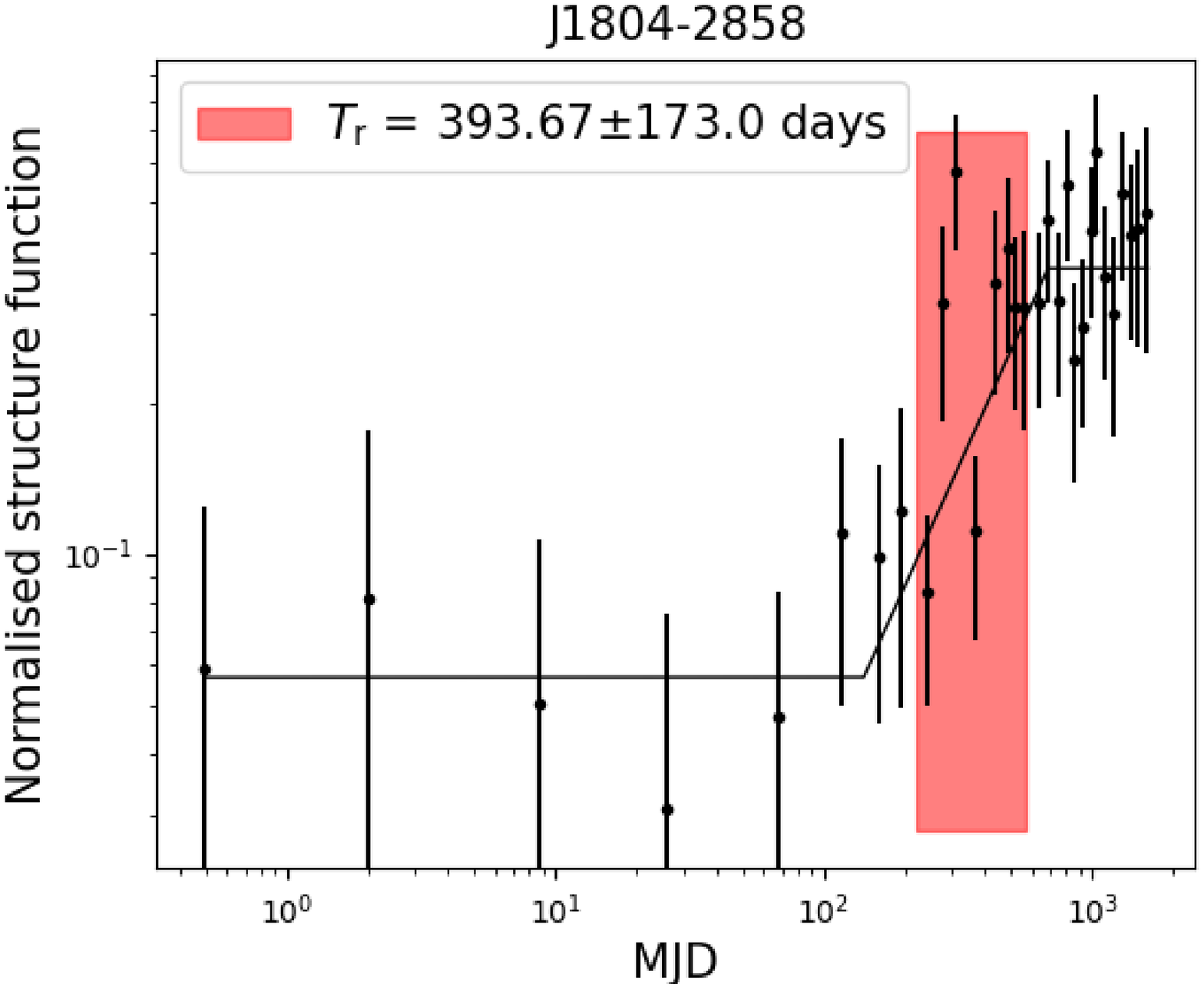}
	\includegraphics[width=5.3cm,height=4.1cm]{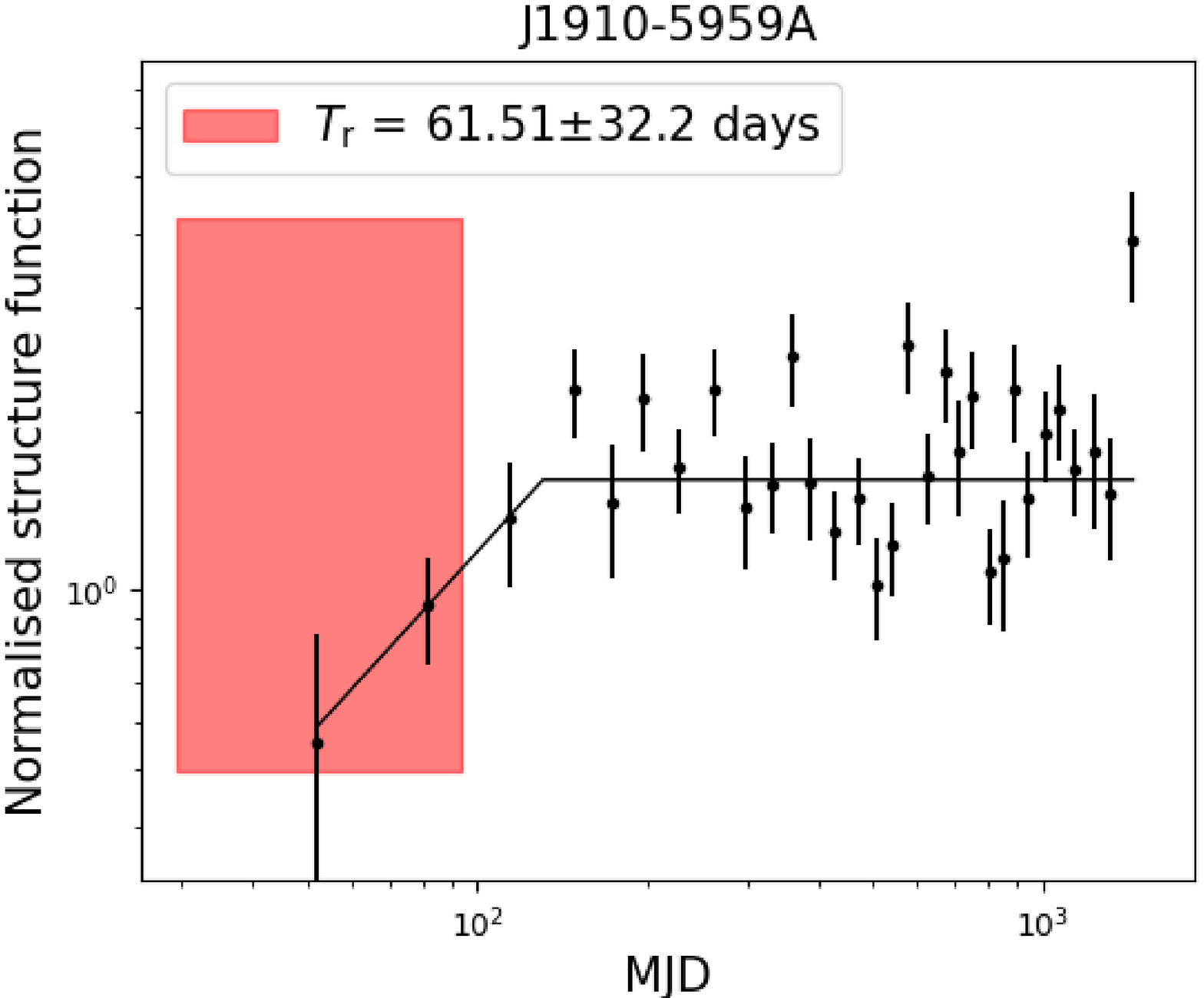}
	\includegraphics[width=5.3cm,height=4.1cm]{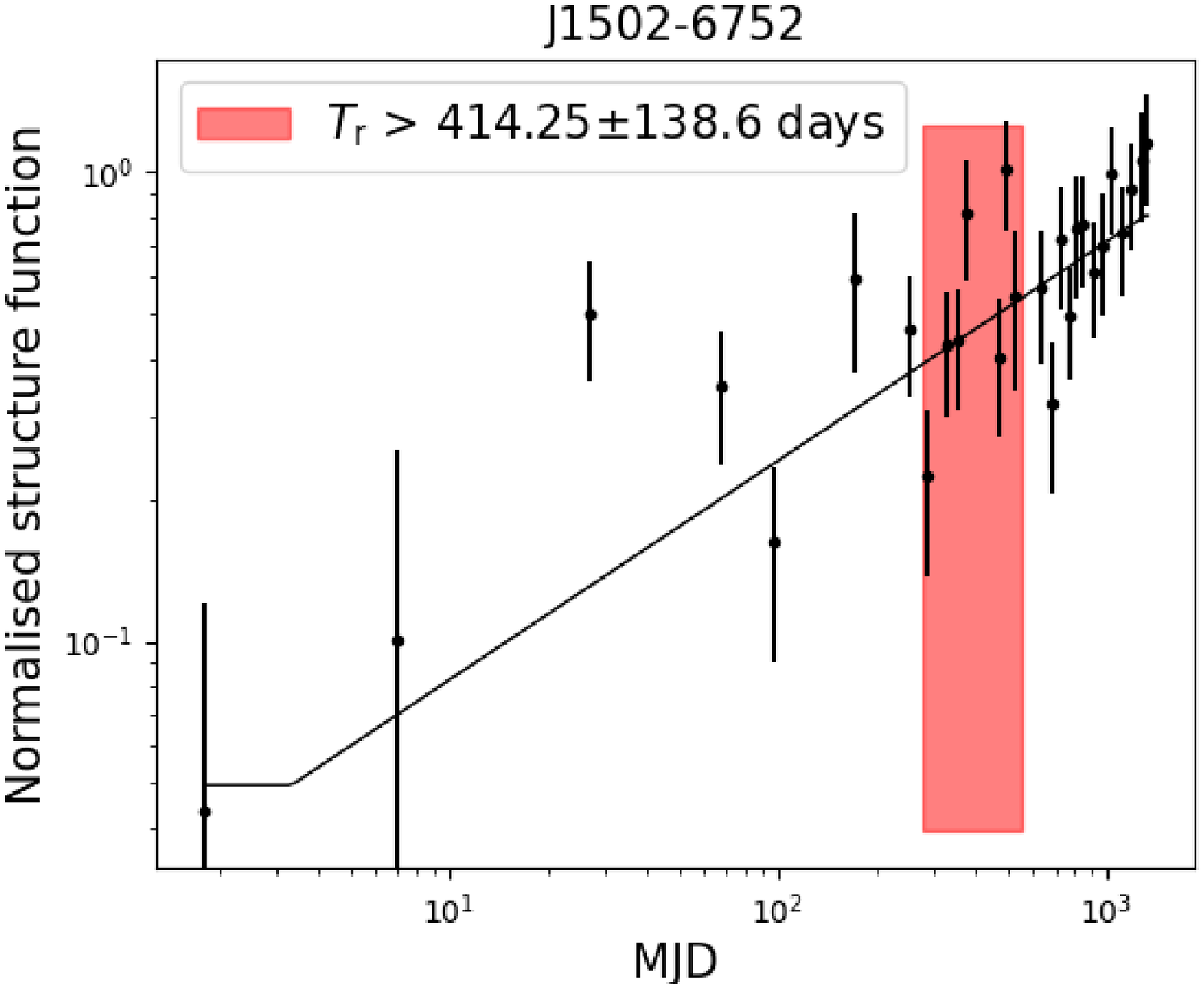}
	\includegraphics[width=5.3cm,height=4.1cm]{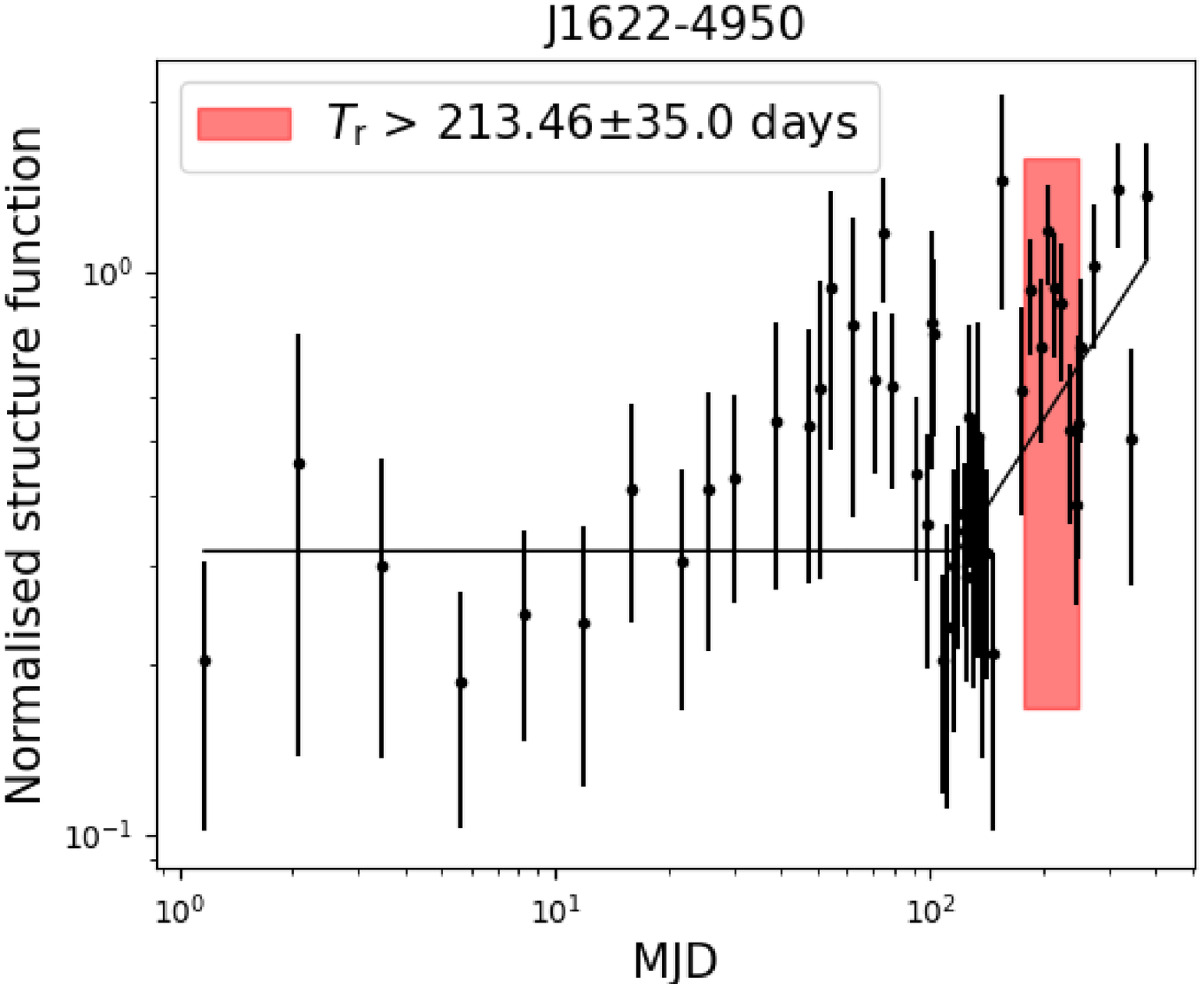}
	\includegraphics[width=5.3cm,height=4.1cm]{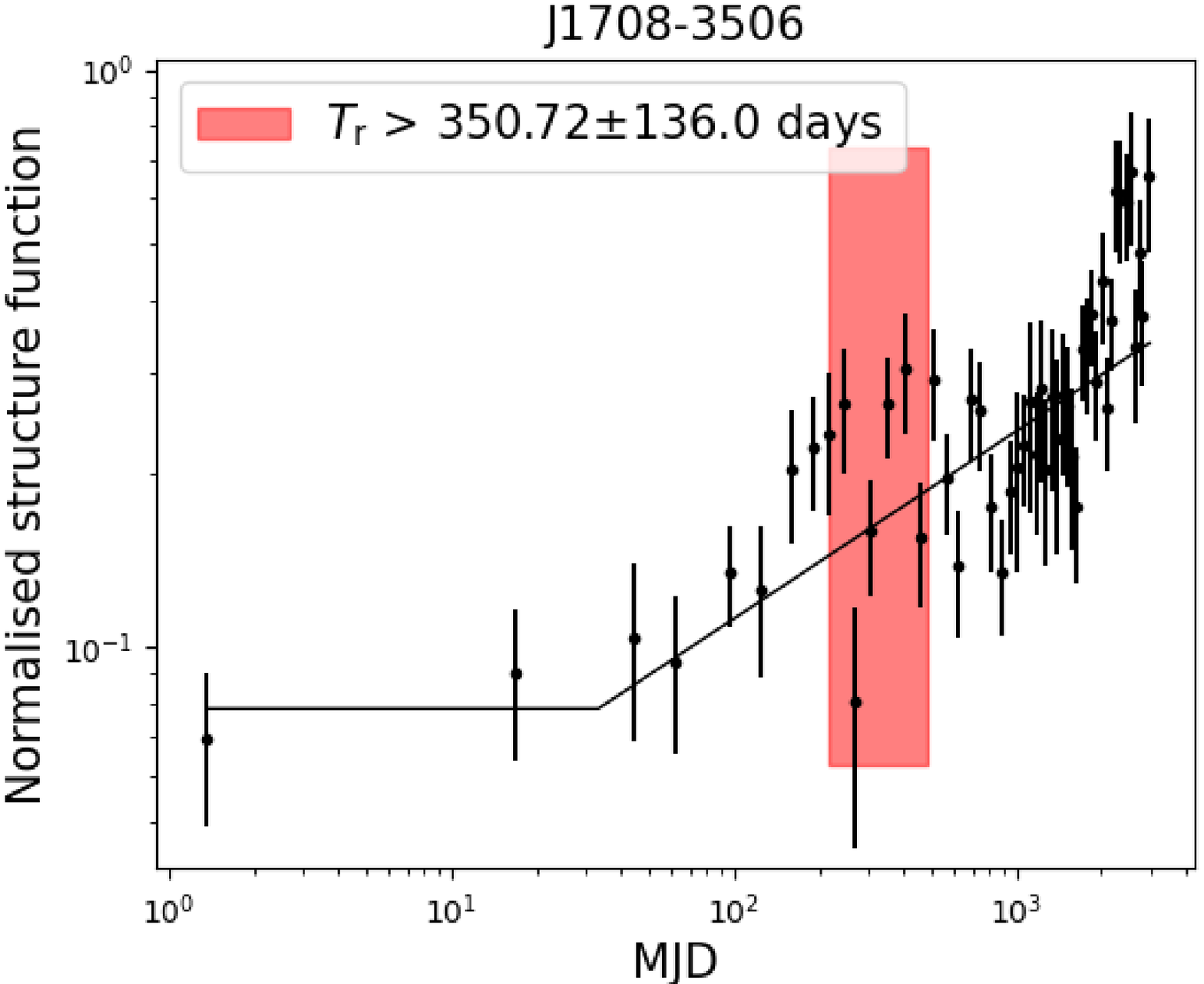}
	\includegraphics[width=5.3cm,height=4.1cm]{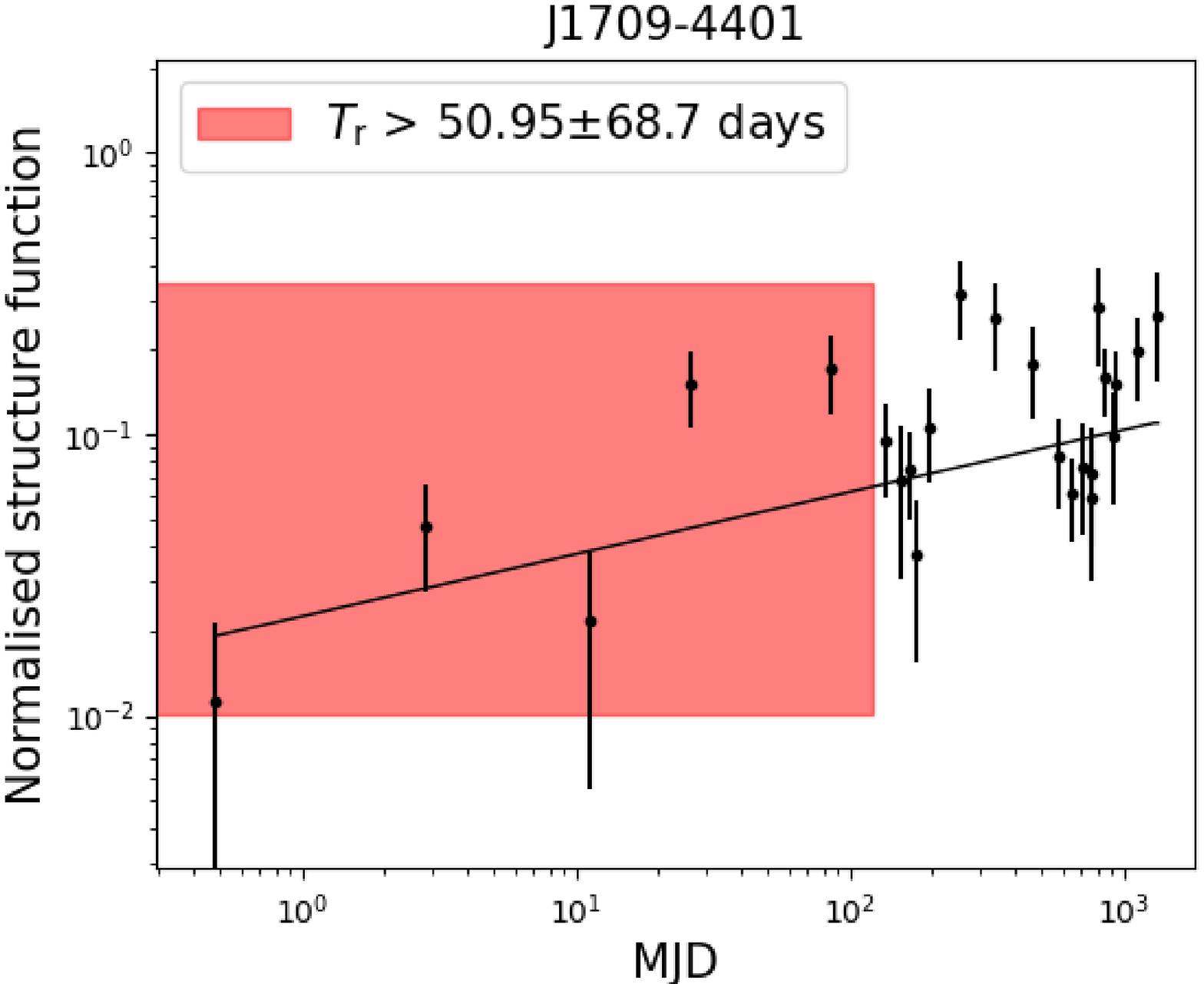}
	\includegraphics[width=5.3cm,height=4.1cm]{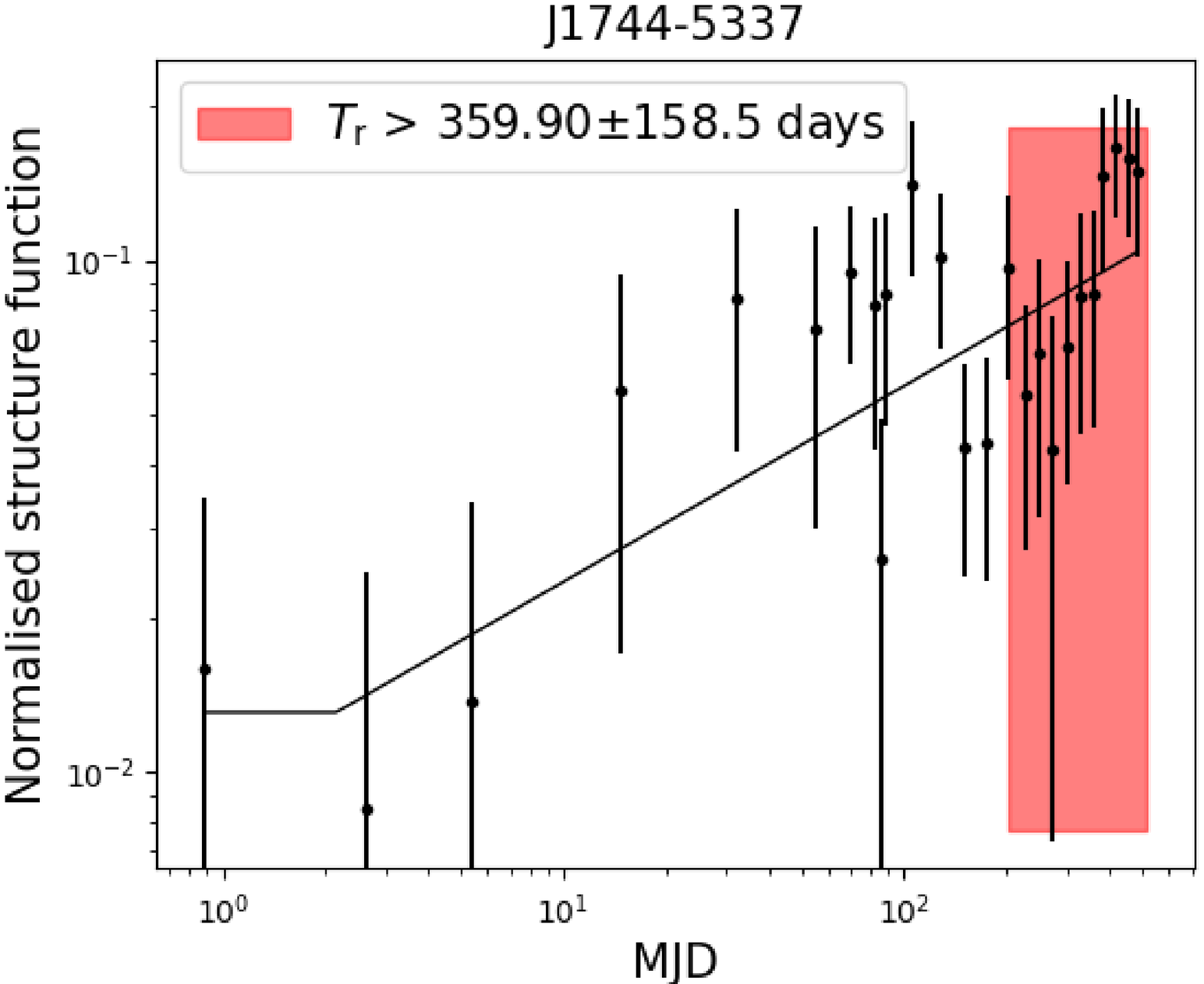}
	\caption{The structure functions diagram of pulsars are classified as `S' and `I'. Those pulsars that belong to the classes `S' and `I' are listed in Table~\ref{tab:S+I}. The black lines indicate the results of the fits to the three regimes  described in Section \ref{structure}. The red area show the measured refractive timescale or lower bound and its uncertainty.}
	\label{sf:S+I}
\end{figure*}

Fig.~\ref{fig:PsrcatFlux} shows a comparison of average flux density measurements from our work with those from the ATNF pulsar catalogue.
We show matches with the catalogue, with 146 pulsars at 1.4 GHz.
Effectively, all our measurements at 1.4 GHz are new ones. As shown in Table~\ref{longtable} and Fig.~\ref{fig:PsrcatFlux}, most pulsars in our sample are weak and have flux density below one mJy. There are a few bright pulsars with flux density over ten mJy. The number of observations for these bright pulsars is small, and they can be studied in detail in the future. 
Most of the data points in Fig.~\ref{fig:PsrcatFlux} are located near the identity line.
{For the first time, uncertainties of average flux density are given for 86 pulsars which account for more than half in Fig.~\ref{fig:PsrcatFlux}.} The green dots show flux density points that deviate from the identity line by at least $5 \sigma$. Thirty-four measurements show significant deviations. Among them, 23 are from publications of different Parkes pulsar surveys. Eleven are from the High Time Resolution Universe Pulsar Survey \citep{2015MNRAS.450.2922N} and five from the Parkes Multibeam Pulsar Survey \citep{2006MNRAS.372..777L}, two from \cite{2007ApJ...656..408J}, one from \cite{2006MNRAS.368..283B}, one from \cite{2015ApJ...810...85C}, one from \cite{2013ApJ...774...93K}, one from \cite{2009MNRAS.395..837K} and one  from \cite{2011MNRAS.414.1292K}.
The differences may not be surprising since their flux densities were estimated using radiometer equations {which only can be served as a rough estimate}. {The other outliers (such as PSR J1634$-$5107 \citep{2015MNRAS.449.1495Y}, J1119$-$6127 \citep{2018MNRAS.480.3584D}, J1809$-$1943 \citep{2006Natur.442..892C})} show nulling or extreme variability in their flux density.
 We notice that seven {outliers} are from \cite{2018MNRAS.474.4629J}. {The flux densities from \cite{2018MNRAS.474.4629J} are calibrated and averaged over multiple observations. Our flux density measurements for these pulsars include more observations than that of \cite{2018MNRAS.474.4629J}, which may indicate  the  flux density variability of these pulsars.}
{In addition to these outliers, the root-mean-square difference relative to the catalogue is 28\%, which reflects the degree of deviation of flux density.} We find that our data are uniformly distributed around the identity line, {which indicates the measurements from the catalogue agree with our estimates in general.}

Fig.~\ref{fig:MJD} shows flux densities of 8 pulsars as a function of MJD as examples.
{The data spans of these pulsars are more extended than ten years, and the observation cadences are relatively uniform. Flux density time-series for each pulsar, as shown in Fig.~\ref{fig:MJD}, are obtained and then used to derive the modulation index and structure function in the following subsections.} A variety of fluctuation behavior is seen in Fig.~\ref{fig:MJD}.  The pulsar flux density (e.g. PSRs J0900$-$3144, J1227$-$6208, and J1801-3210) is stable over 10 yr time span. {Variations of the
flux density for these pulsars are comparable with the
uncertainty.} 
The flux density of some other pulsars (e.g. PSRs J1105$-$6107, J1125$-$6014, and J1224$-$6407) exhibits significant fluctuation with different characteristics timescale.

\subsection{Modulation index}

A proper parameter to characterize the variability can be the modulation index, $ m ~=~ \frac{\sigma_{S}}{\overline{S}}$, which originates from a combination of different effects, including the radiometer noise, the internal intensity  fluctuation of the pulsar, and the diffractive and refractive scintillation. We have quantified these effects following the scheme of \cite{2021MNRAS.501.4490K}. {In the flux density measurements, the contribution of radiometer noise can be estimated as $m_{\rm{n}} ~=~ \frac{\overline{e_{\rm{t}}}}{\overline{S}}$. `jitter noise' is caused by the pulse-to-pulse variation, and its contribution to modulation index  is estimated as: $m_{\rm j} ~=~ \sqrt{\frac{1}{N_{\rm pulse}}} ~=~ \sqrt{\frac{P}{t_{\rm obs}}}$. 
The modulation index after correcting for jitter and measurement noise is estimated as
$m_{\rm corr} ~=~ \sqrt{m^{2}-m^{2}_{\rm n}-m^{2}_{\rm j}}$.}

{We can calculate the expected refractive scintillation modulation index, $m_{\rm r}$, and diffraction scintillation modulation index, $m_{\rm d}$. 
{It is necessary to know the diffractive scintillation timescale, $t_{\rm{d}}$ and the scintillation bandwidth $\Delta\nu_{\rm d}$.
\cite{1998MNRAS.297..108J} estimates the value of $t_{\rm d}$ as 
\begin{eqnarray}
\label{eq;td}
t_{\rm{d}} ~=~ \frac{3.85\times10^4 \,\, \sqrt{D \Delta\nu_{\rm{d}}}}{\nu \,\, V_{\rm{iss}}},
\end{eqnarray}
where the transverse velocity, $V_{\rm{iss}}$ (in km\,s$^{-1}$), and pulsar distance, $D$ (in kpc), are taken from the ATNF Pulsar Catalogue. $\nu$ (in GHz) is the central frequency of the observation. 
The value of $\Delta\nu_{\rm d}$ (in MHz) can be calculated by
\begin{eqnarray}
    \label{eq;bandwidth}
    \Delta\nu_{\rm d} ~=~ \frac{1.16}{2\pi \tau_{\rm s}},
\end{eqnarray}
}
Thus, we need to estimate the scattering time, $\tau_{\rm s} ~=~ 1.2 \times 10^{-5} \,\, {\rm DM}^{2.2}(1.0 + 0.00194 {\rm DM}^2)$ \citep{2015ApJ...804...23K}. Assuming a uniform medium with a Kolmogorov wave number spectrum~\citep{1998ApJ...507..846C}, the modulation index caused by diffractive scintillation, $m_{\rm{d}}$, can be estimated as $m_{\rm{d}} ~=~ \sqrt{5 \,\, \frac{\Delta\nu_{\rm{d}}}{\Delta\nu} \,\, \frac{t_{\rm{d}}}{t_{\rm{obs}}}}$. For the Kolmogorov spectrum with uniform scattering medium and a small inner scale, $m_{\rm r}$ can be roughly estimated as $m_{\rm{r}} ~=~ 1.1 \biggl{(}\frac{\Delta\nu_{\rm{d}}}{\nu}\biggr{)}^{0.17}$~\citep{2000ApJ...539..300S}.}

Columns 8$-$13 of Table~\ref{longtable} list \emph{$m$}, \emph{$m_{\rm n}$}, \emph{$m_{\rm j}$}, \emph{$m_{\rm corr}$}, \emph{$m_{\rm d}$}, and \emph{$m_{\rm r}$}, respectively.
{Modulation indexes for 95 pulsars are obtained. As shown in Fig.~\ref{fig:MJD}, even for pulsars with a small modulation index, the difference between the minimum and maximum flux density is still significant. It is well-known pulsar flux varies in time and frequency, and we recommend that the modulation index of this work be included in the next version of the ATNF pulsar catalogue to reflect the variability of pulsar flux density.}
Note that the estimations {\emph{$m_{\rm d}$} and \emph{$m_{\rm r}$}} are based on a {small and thin scattering screen model} of the free electron distribution in the ISM. The uncertainties in estimates of scattering time, pulsar velocities, and distances are not well known and can be very large. Therefore,  {\emph{$m_{\rm d}$} and \emph{$m_{\rm r}$}}  estimates can be served as a rough estimate for any particular pulsar. However, a comparison of the sizes of the {\emph{$m_{\rm n}$}, \emph{$m_{\rm j}$}, \emph{$m_{\rm corr}$}, \emph{$m_{\rm d}$}, and \emph{$m_{\rm r}$} allows a preliminary view to be made of the possible dominant factor of the modulation index \emph{$m$} {\citep{2021MNRAS.501.4490K}}}.

\subsection{Structure functions}
\label{structure}
Structure function, $D(\tau)$, has been widely used in the study of pulsar scintillation {\citep{1985ApJ...296...46S} as follows:
\begin{equation}
    D(\tau) = \langle[S(t+\tau)-S(t)]^{2}\rangle,
\end{equation}
where the angle brackets represent an ensemble average and $\tau$ is the time lag.} Here, we used structure functions to analyze the flux density variation in more detail. The advantage of structure function is that it is not affected by the non-uniform sampling of time series and can be used to measure refractive timescales. Ideally, the expected structure function has three different regimes: {(1)flat structure function with short lag (2) rising structure function with medium lag, and (3) flat structure function with long lag. In the first regime, the amplitude of the structure function is the result of a combination of the variability in the pulsar itself, the uncertainty in the measured flux density, the RISS wave spectrum shorter than the mean interval between observations for a specific, and unquenched DISS. The rising regime of the structure function is represented by the power-law function with a logarithmic slope $\gamma$. The time required to reach saturation is correlated with the refraction timescale, and the level of the structure function for those pulsars at long lag is determined by the RISS characteristic of ISM.}

We calculated the structure functions for pulsars observed more than 30 times and classified them according to the scheme of \cite{2021MNRAS.501.4490K}. Structure functions showing all three regimes or a rising slope with saturation are classified as `S'.  Structure functions that show the continuous increase are classified as `I' and imply that the maximum lag is shorter than the refractive timescale. 
Pulsars with a flat structure function are classified as `F', and these pulsars account for the majority.
These flat structure functions could result from {different reasons, including the intrinsic intensity variability, the radiometer noise, and the diffractive and refractive effects}. We categorized them by the factors that dominate the modulation index. 
{
\begin{enumerate}
    \item Class `F-N': weak pulsars and/or measurement noise
    \item Class `F-J': jitter noise
    \item Class `F-R': refractive scintillation, but our shortest lag is longer than the refraction time
    \item Class `F-D': diffractive scintillation 
    \item Class `F-DR': combination of diffractive and refractive scintillation
\end{enumerate}}

Finally, we created an `X' Class with significantly abnormal modulation index for pulsars. The last column of Table~\ref{longtable} lists the classification.


\begin{table}
    \centering
    \caption{Refractive parameters. Uncertainties are given in parenthesis. {The noise and saturation are the levels of the structure function for small and large lags, respectively.}} 
	\label{tab;ref}
\begin{tabular}{lrrrr}
\hline
\multicolumn{1}{c}{Name} & \multicolumn{1}{c}{$T_{\rm r}$} &  \multicolumn{1}{c}{Noise} &  \multicolumn{1}{c}{Slope}  & \multicolumn{1}{c}{Saturation}  \\
     & \multicolumn{1}{c}{[d]}    &       &  \multicolumn{1}{c}{[d$^{-1}$]} & \\ 
\hline
\multicolumn{5}{c}{Class `S'} \\
    J0900$-$3144 & 52(18) &       & 0.8(3) & 0.0223(7) \\
    J1101$-$6424 & 24(34) & 0.012(5) & 0.4(4) & 0.035(3) \\
    J1105$-$6107 & 55(35) &       & 0.5(2) & 0.99(1) \\
    J1125$-$5825 & 1(4)  &       & 0.2(1) & 0.73(2) \\
    J1435$-$6100 & 4(10) & 0.09(2) & 0.2(2) & 0.148(6) \\
    J1545$-$4550 & 6(16) &       & 0.22(8) & 0.060(2) \\
    J1753$-$2240 & 692(140) & 0.035(6) & 1.2(3) & 0.22(5) \\
    J1802$-$2124 & 50(31) &       & 0.5(2) & 0.028(1) \\
    J1804$-$2858 & 394(170) & 0.06(2) & 1.0(4) & 0.39(5) \\
    J1910$-$5959A & 62(32) &       & 0.8(4) & 1.53(4) \\
\multicolumn{5}{c}{Class `I'}\\
    J1502$-$6752 & $>$414(140) & 0.18(4) & 0.6(2) &  \\
    J1622$-$4950 & $>$213(35) & 0.32(2) & 1.0(3) &  \\
    J1708$-$3506 & $>$351(140) & 0.08(1) & 0.33(6) &  \\
    J1709$-$4401 & $>$51(69) &       & 0.3(1) &  \\
    J1744$-$5337 & $>$360(160) & 0.04(1) & 1.4(8) &  \\
\hline
\end{tabular}
\label{tab:S+I}
\end{table}
\begin{figure}
	\centering
	\includegraphics[width=\columnwidth]{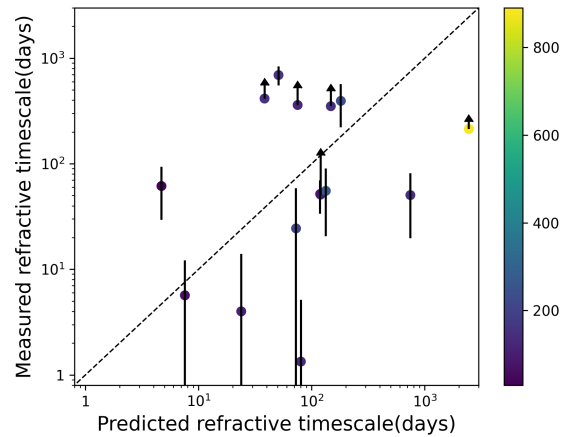}
	\caption{Refractive timescales for prediction and measurement of pulsars classified as `S' and `I'. The predicted refractive timescales are estimated using $t_{\rm{r}} ~=~ \frac{4}{\pi} \,\, \frac{\nu}{\Delta\nu_{\rm{d}}} \,\,\, t_{\rm{d}}$ \citep{1990ApJ...352..207S}. It is worth noting that the refractive timescales by this equation may be significantly different from the actual value and can only be indicative. The color bar indicates pulsar DMs, in ${\rm cm}^{-3} \rm pc$.}
	\label{fig:pre-mea}
\end{figure}

\subsection{Refractive scintillation parameters}
{A typical structure function exhibits three distinct regimes: a noise regime at small lags, a structured regime characterized by a rising slope in a log-log plot, and a saturation regime where the structure function flattens out. }
For structure functions exhibiting a rising slope and/or saturation {(Classes `I' and `S')}, we determined the slope of the rising state in logarithmic coordinates and levels of saturation and noise by fitting a combination of a single power-law and constant levels to the structural functions. {In Fig.~\ref{sf:S+I}, we present the structure functions and fitting results of 10 class `S' pulsars and 5 class `I' pulsars. The solid black lines show the fitting results of noise, rising, and saturation of the structure functions. Some pulsars of class `S' and class `I' (such as PSR J0900$-$3144 and J1708$-$4406) do not exhibit noise regions of small lag, and all pulsars of class `I' do not show saturation regions.}{For class `F' pulsars, the RISS parameter can not be estimated.} 

As mentioned above, the structure functions are piece-wise functions with three regions, and the demarcation point of the function is unknown. Therefore,  it is not easy to determine the best fitting parameters and their uncertainties with the least square fitting.
We used the Bayesian inference package {\sc bilby}~\citep{2019ApJS..241...27A} to fit the data and estimate the uncertainties of parameters. 
We assume that each parameter has a Gaussian likelihood function of a prior. 
The {\sc dynesty} sampling algorithm in {\sc Bilby} is used to sample the posterior distribution of parameters~\citep{2020MNRAS.493.3132S}.
The mean of the posterior distribution for each parameter deviating from the median at 16 percent and 84 percent were estimated as its uncertainty.
We increase the uncertainty of the estimated structure function and refit the model till the $\chi$-squared of the fitting decreases to less than 5, which yields a reliable estimate of the scintillation parameter.

Measured RISS parameters {from Fig.~\ref{sf:S+I}} of 15 pulsars of classes `I' and `S' are listed in Table~\ref{tab:S+I}. 
The refractive timescale $T_{r}$ (Column 2) is calculated as the lag for the structure function to reach half its saturated value. 
We calculated the lag for the structure function to reach half its saturated value as the refractive timescale $T_{r}$ (Column 2).
Only the lower bound is given for class `I' pulsars.  Column 4 shows the slope of the rising area. 
The structure function slope can reflect the distribution of ionised medium along the line of sight \citep{1986MNRAS.220...19R}.
The predicted slope is ~2.0 for a Kolmogorov spectrum and a thin screen. 
When the ionized medium is uniformly distributed along the line of sight, the slope is close to $\beta - 3$. Measured slopes of 15 pulsars are presented in Table~\ref{tab;ref}. Most of our samples show slopes in the range of 0.4–1.0. {The slopes of two pulsars are in the range of 1.2-1.4.} No pulsar has a slope close to 2. 
{The expected
slope for a Kolmogorov spectrum and an extended distribution of scattering material is 0.67 which is consistent with most the measured slopes.
Pulsars with slopes close to 2 tend to cluster at low DMs \citep{2021MNRAS.501.4490K}. However, most of pulsars with measured slopes have medium or high DM. 
In the future, possibly we need a new model of distribution of ISM or a new high sensitive instrument for next generation observing.
}
The Column 5 represents the saturated level of structure function of class `S' pulsars. A wide range of fluctuation characteristics is evident in Table~\ref{tab:S+I} and Fig.~\ref{sf:S+I}. Saturated structure function are evident with a wide range of saturation levels, all the way from 0.022 for PSR J0900$-$3144 to 1.53 for PSR J1910$-$5959A. Note that not all 15 pulsars exhibit obvious noise levels at small lags, and Column 3 only presents those that can be measured.
{We compare the measured value of refractive time-scale with the predicted value obtained {using $t_{\rm{r}} ~=~ \frac{4}{\pi} \,\, \frac{\nu}{\Delta\nu_{\rm{d}}} \,\,\, t_{\rm{d}}$}, in Fig.~\ref{fig:pre-mea}. Given the uncertainty of the pulsar's distances and transverse velocities, a large scatter can be seen, which is not surprising.}


\section{DISCUSSION}
\label{DISCUSSION}
{We provide {calibrated} average flux density measurements for 151 pulsars, modulation index for 95 pulsars, and structure function for 54 pulsars. These pulsars have a wide DM, distance, and period range from $\sim$ 3 to $\sim$ 1000 $\rm{cm^{-3} pc}$, $\sim$ 0.2 to $\sim$ 25 kpc, and $\sim$ 1.5 ms to $\sim$ 8.5 s, respectively, which makes them valuable for studies of pulsar spectra and RISS. This section discusses the relationship between modulation index and scattering strength, outlier pulsars, structure function of Class `F', and the relationship between modulation index and pulsar distribution.}

\subsection{Scattering strength and modulation index}
\begin{figure*}
	\centering
	\includegraphics[width=16cm]{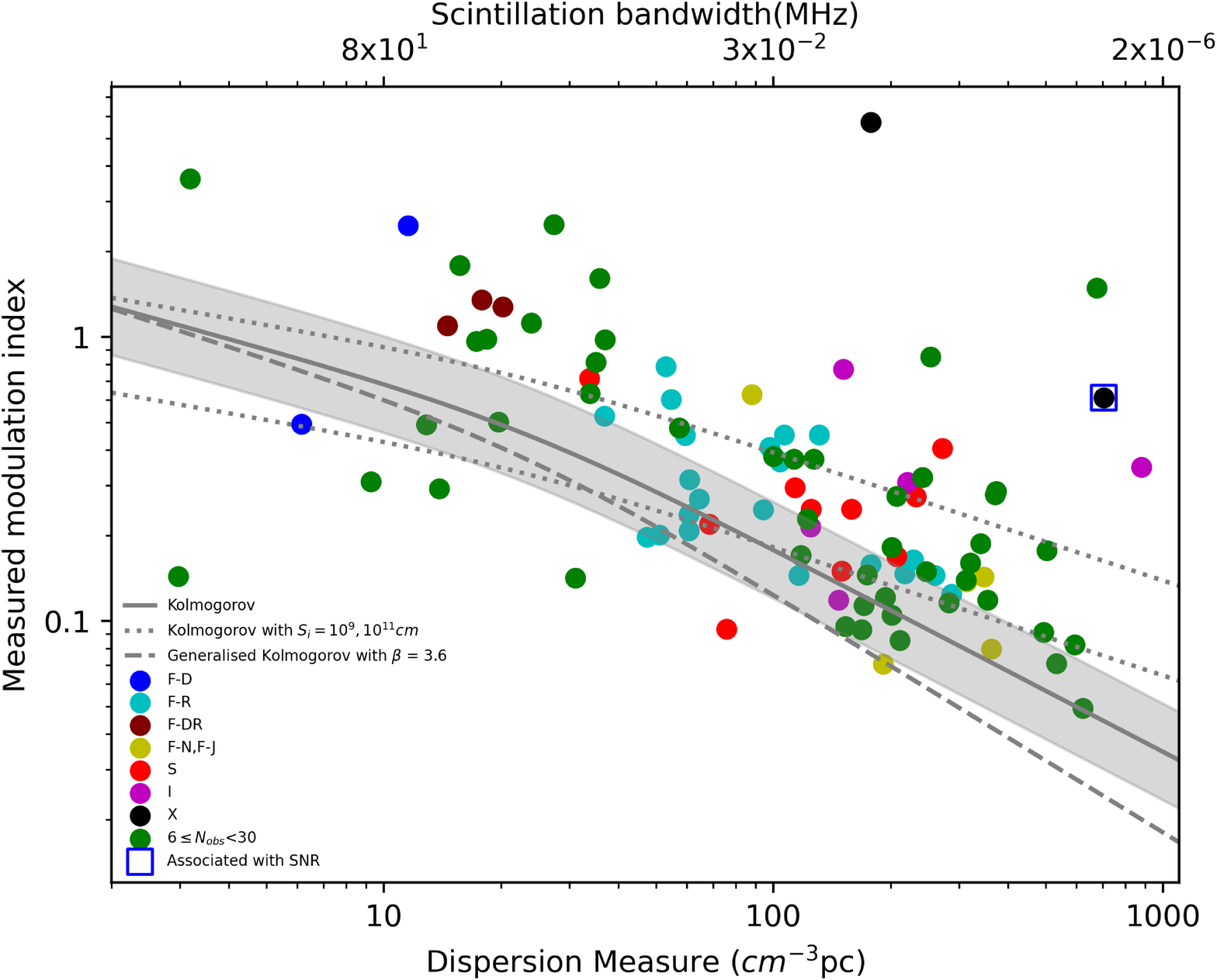}
	\caption{
	The relationship between the modulation index of 95 pulsars and DM. Different colors represent different pulsar classifications in the text.
	Pulsars that match the number of observations $6 \leq N_{\rm obs} < 30$ are indicated by green dots. The top axis shows the estimated scintillation bandwidth, $\Delta\nu_{\rm d}$, for a given DM \citep{2000ApJ...539..300S}. 
	The solid line indicates the function between the refractive modulation index predicted by Kolmogorov spectrum and DM.
	We present the region within an order of magnitude around the estimated scattering timescale as the shaded area. 
	The dotted lines at the top and bottom show the prediction of the Kolmogorov model, and the `inner scale' is ${10}^{9}$cm and ${10}^{11}$cm, respectively. 
	The dashed line represents the prediction curve of the generalized Kolmogorov spectrum with index $\beta$ = 3.6.
	We refer to the ATNF Pulsar catalogue to mark pulsars that may be associated with SNRs regions with squares.
	}
	\label{fig:dm_m}
\end{figure*}

Fig.~\ref{fig:dm_m} presents the modulation index corrected for measurement noise correction, $\sqrt{m^{2}-m^{2}_{\rm{n}}}$, as a function of DM. The Pulsars are colored based on the classification scheme mentioned above. {For pulsars with few observations, we also get their modulation indexes. Considering that these pulsars' time series points are relatively small and the obtained modulation index error is rather large, we did not classify these pulsars. 
We can continue to observe these pulsars and obtain their accurate modulation index in the future. {More than 1400 pulsars have been observed using Parkes telescope. Only a small part of them have been monitored for years. In principle, all the modulation index can be obtained.}
A color distribution can be seen in Fig.~\ref{fig:dm_m}. The modulation index is mainly dominated by diffraction scintillation, at low DMs. 
With the increase of DM, the scattering effect becomes more robust, and refractive scintillation gradually dominates.
When the DMs are over 100 $100$\,cm$^{-3}$\,pc, the modulation indices of most pulsars appear to be smaller than $\sim0.4$, which is consistent with results of ~\cite{2020MNRAS.493.3132S}, which suggest that the pulsar flux density is constitutionally stable.}
{The dispersion of the modulation index is large for a given DM. 
Our sample of pulsars has 61 pulsars that DM is greater than 100\,cm$^{-3}$\,pc, covering a wider range of DM.
This sample is very valuable for studying refractive scintillation, and the influence of DISS on the variability of pulsar flux density can be ignored.}

The modulation index of RISS enables us to detect the intensity of scintillation and confine the turbulence spectrum.
Some previous researches on RISS compared the diffraction scintillation bandwidth and refractive modulation index, and limited the internal scale and power-law density inhomogeneity spectrum~\citep[e.g.][]{2000ApJ...539..300S,1987ApJ...315..666C}.
The receiver used in our work can not get the measurement of DISS bandwidth very well, in particular for pulsars with high DM, we can use the Parkes Ultra-Wideband Low receiver~\citep[UWL;][]{2020PASA...37...20K} in the future.
In our work, We have compared the observed distribution of modulation indices  with theoretical predictions.

{The solid line in Fig.~\ref{fig:dm_m} represents the expected RISS modulation index for a Kolmogorov spectrum as a function of DM. 
The shaded areas are areas within an order of magnitude around the average trend.
We also plotted the prediction of a generalised Kolmogorov spectrum of an index of $\beta=3.60$ with dashed line~\citep{2000ApJ...539..300S} and a Kolmogorov spectrum of different inner scales with dotted lines~\citep{1993ApJ...403..183G}. 
Even considering the great uncertainty of using DM to estimate the scattering time (shadow region), the simple Kolmogorov model can not account for the large scattering.
Some pulsars exhibit much larger variability than predicted using a simple Kolmogorov model, which could be explained by the Kolmogorov model with inner scales  10$^9$ cm. 
A plausible explanation for this result is the turbulence of ISM varies greatly along different lines of sight.
Despite the large scatter, there is a moderate inverse correlation between DM and the modulation index in a wide DM range with a correlation coefficient of -0.47. }

\subsection{Outlier pulsars} \label{outlier}

We check the significant outliers in the table and figures and get pulsars whose modulation index is clearly different from the expectations. There are two pulsars classified as `X' in Table~\ref{longtable}.

PSR J1119$-$6127 is a high-B, rotation-powered radio pulsar. Magnetar-like X-ray bursts from the pulsar were detected in 2016. {Previous research \citep{2018MNRAS.480.3584D} showed that the radio pulses disappeared after the X-ray bursts and reappeared about two weeks later, with a five-fold increase in the 1.4 GHz flux density. Then the flux density drops below the normal flux density and slowly returns to normal. The pulsar’s integrated profile underwent dramatic and short-term changes in total intensity, which strongly affects the modulation index.}

PSR J1809$-$1943, an anomalous X-ray pulsar, was the first magnetar found to be emitting radio pulsars after a strong high-energy outburst \citep{2003IAUC.8190....2G}. {Previous studies \citep{2006Natur.442..892C} have shown that there was no evidence of radio emissions prior to the 2003 X-ray burst (unlike normal pulsars, which emit radio pulses all the time), and that the flux varies from day to day.}
This results in the predicted modulation index of DM much lower than the observed modulation index.

\subsection{Structure function of Class `F'}
{We derive the structure function of 54 pulsars. 39 structure functions are categorized as `F'. Most of the structure functions from \cite{2021MNRAS.501.4490K} also belong to `F' class. RISS parameters can not be estimated for these pulsars. In fact, the RISS parameters have been determined for less than 100 pulsars which only account for a very small fraction of known radio pulsars. The modulation index of flat structure functions can be caused by many reasons. According to the classification in Sec 3.3, the RISS parameter could be possible measured for some class `F' pulsars in the future with different observing strategies:
\begin{enumerate}
\item  observations using longer integration or telescope with higher sensitivity to improve the signal to noise ratio of observation for Class `F-N'.
\item  observations using a telescope with higher sensitivity for Class `F-N'.
\item observations with higher cadence for Class `F-R' .
\end{enumerate}
{In addition, the data analysis methods could possibly refine with high quality measurement of polarisation. It has well known to have the dependence on the distance and ISM to pulsar. Recent works as \cite{2021MNRAS.504..228S} found general trends in the pulse profiles including decreasing fractional linear polarization and increasing degree of circular polarisation with increasing frequency. Either linear or circular polarisation is possibly valuable to distinguish available pulsars from others.}
The modulation index of these pulsars will not be dominated by measurement noise, jitter noise, and RISS using such observing strategies. And then, we could possibly measure the RISS parameters from the structure function.}



\subsection{Modulation index and pulsar distribution}
\begin{figure*}
	\centering
	\includegraphics[width=15cm]{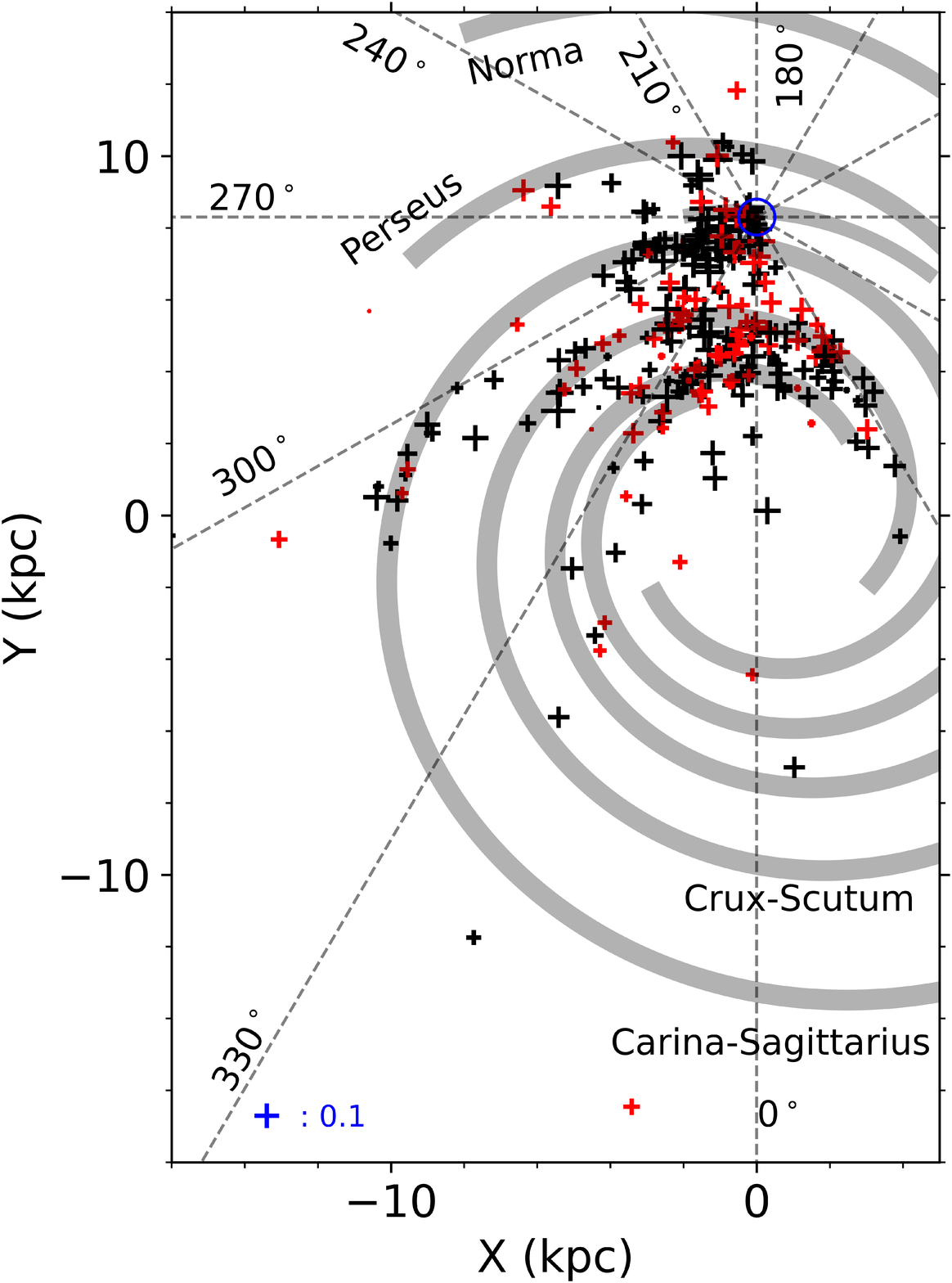}
	\caption{The distribution of pulsars with measured modulation index on the galactic plane. The blue circle is the position of the solar system. The spiral arm parameters of the Galaxy are taken from \citet{2014A&A...569A.125H}. The size of the plus sign is proportional to the deviation between the predicted and measured values of the modulation index {and displayed with logarithmic scale. The blue plus on lower left indicates the size of deviation is 0.1. The black and red pluses
represent pulsars whose predicted modulation index is smaller or
larger than the measured value, respectively.}}
	\label{fig:galaxy}
\end{figure*}
\begin{figure}
	\centering
	\includegraphics[width=\columnwidth]{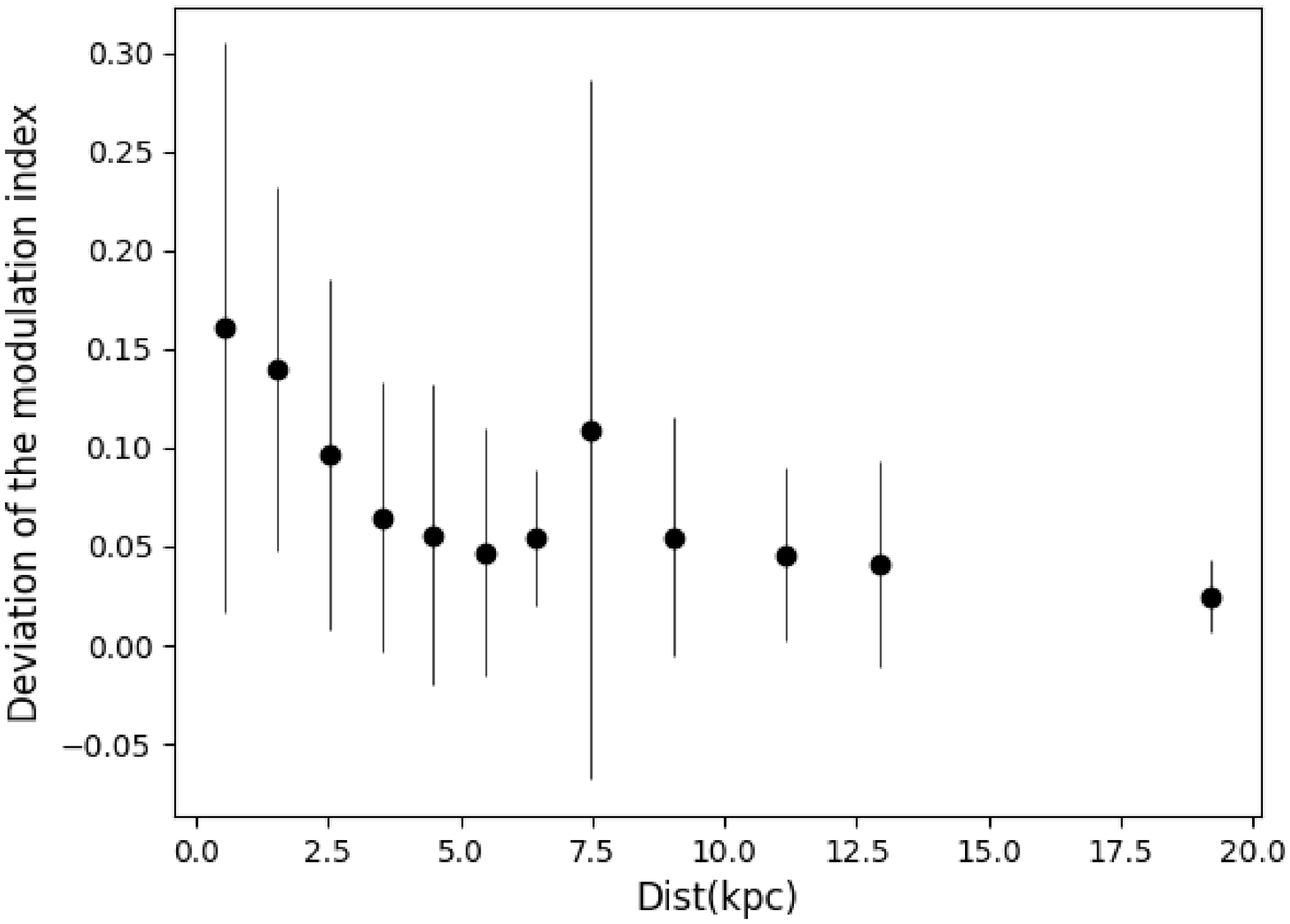}
	\caption{Deviation of modulation index as a function of pulsar distance.}
	\label{fig:pulsars_distance_distribution}
\end{figure}

Fig.~\ref{fig:galaxy} shows the distribution of pulsar with the deviation between the measured modulation index and prediction for a Kolmogorov spectrum from this paper and \cite{2021MNRAS.501.4490K} on the galactic plane. The pulsar distance is estimated from the YWM16 model. We only include pulsars with DM {$>$} 30, which indicates the modulation indexes of these pulsars are not dominated by DISS. The blue circle indicates the position of the solar system. The black and red pluses represent pulsars whose predicted  modulation index is smaller or larger than the measured value. {The size of the plus sign is proportional to the deviation between the predicted and measured values of the modulation index. We use logarithmic scale for the size of black and red plus. The blue plus on the lower left indicates the size of deviation is 0.1. The deviations range from 0.0005 to 0.64.} It is clear that most pulsar modulation index measurements are larger than predicted. In some directions (0$^\circ$ $\sim$ 30$^\circ$), the modulation indexes are more consistent with the prediction than that in other directions. Pulsars within $\sim$ 2 kpc from the solar system show a relatively large modulation index deviation, which may be caused by diffractive scintillation. The deviations of the modulation index as a function of distance are shown in Fig.~\ref{fig:pulsars_distance_distribution}. The deviations of the modulation index are averaged over 1 kpc for pulsar within 10 kpc. And for pulsar $>$ 10 kpc, the modulation index is averaged according to the number of pulsars. The standard deviation is shown as an error bar in Fig.~\ref{fig:pulsars_distance_distribution}. For pulsar within 5 kpc, the averaged deviation of the modulation index decrease with increasing distance. Then the average deviation remains steady.  Nearby pulsars with low DMs may be more likely affected by diffractive scintillation, and  the deviation is larger.

\section{CONCLUSIONS}
\label{CONCLUSIONS}
We measured the average flux density and its uncertainty of 151 radio pulsars at 1.4 GHz, representing a large pulsars sample for flux density variability studies. {We recommend our results be included in the next version of the ATNF pulsar catalogue. The modulation index and structure function for 95 and 54 pulsars are obtained. We recommend that the modulation index be included in the next version of the ATNF pulsar catalogue to manifest the flux density variation of pulsars. The RISS parameters are estimated for 15 pulsars. The structure functions of most pulsar are flat, consistent with \cite{2021MNRAS.501.4490K}. RISS parameters have been only estimated for a very small fraction of known radio pulsars. We suggest different strategies to measure the RISS parameters in the future for pulsars with flat structure function.} 
These pulsar samples have a wide DM range, and we find that the DM is mildly correlated with the modulation index measurements, but with large scattering.
This implies that the properties of ISM in different lines of sight in the Galaxy are very different. {The data from the UWL receiver will be very helpful to measure the scintillation bandwidth of pulsars. Combining UWL data with our results, we could improve our understanding of RISS and investigate different turbulence models.} 
We will also be able to detect local ISM inhomogeneities and structures along a large number of lines of sight.

\section*{Acknowledgements}

The Parkes radio telescope is part of the Australia Telescope National Facility, which the Australian Government funds for operation as a National Facility managed by CSIRO.
This paper includes archived data obtained through the CSIRO Data Access Portal (http://data.csiro.au). This work is supported by the Zhejiang Provincial Natural Science Foundation of China under Grant (No.LY23A030001), the Major Science and Technology Program of Xinjiang Uygur Autonomous Region (No.2022A03013-4), the Natural Science Foundation of Xinjiiang Uygur Autonomous Region (No.2022D01D85), the National SKA Program of China (No. 2020SKA0120100), the National Natural Science Foundation of China (No.12041304), the CAS Jianzhihua project, the 201* Project of Xinjiang Uygur Autonomous Region of China for Flexibly Fetching in Upscale Talents. The authors thank Dr.Hobbs G. for his valuable suggestions and comments.

\section*{Data Availability}

The observations from the Parkes radio telescope are publicly available from https://data.csiro.au/ after an 18 month embargo period. Note that all original data used in this paper are out of this embargo period and are available.



\bibliographystyle{mnras}
\bibliography{mnras_template} 



\appendix
\onecolumn
\section{TABLE OF OBSERVED PULSARS}
{

\setlength{\tabcolsep}{3pt}
\begin{longtable}{lcccccccccccccc}
\caption{Table of all 151 pulsars in this sample. T is the time span of the data set. $N_{\rm obs}$ is the total number of flux density measurements. $m$,$m_{\rm n}$,$m_{\rm j}$,$m_{\rm d}$, and $m_{\rm r}$ are measured modulation index and contributions from the measurement noise, pulse jitter, DISS, and RISS. $m_{\rm corr}$ is the jitter and noise corrected modulation index. $T_{\rm r}$ is the estimated RISS time scale.}
\label{longtable}\\
\hline
\hline
\multicolumn{1}{c}{Name}& $P_{0}$ & DM & Dist & T & $N_{\rm obs}$ & \(\overline{S}\)  & $m$ 
& $m_{\rm n}$ & $m_{\rm j}$ & $m_{\rm corr}$ & $m_{\rm d}$ & $m_{\rm r}$ & $T_{\rm r}$ & Class\\
 & (s) & ($\rm{cm^{-3} pc}$) & (kpc) & (d) &  & (mJy)  &  & &  &  &  &  & (d)   \\
\hline
\endfirsthead
\caption{Continued.} \\
\hline
\multicolumn{1}{c}{Name}& $P_{0}$ & DM & Dist & T & $N_{\rm obs}$ & \(\overline{S}\)  & $m$ 
& $m_{\rm n}$ & $m_{\rm j}$ & $m_{\rm corr}$ & $m_{\rm d}$ & $m_{\rm r}$ & $T_{\rm r}$ & Class\\
 & (s) & ($\rm{cm^{-3} pc}$) & (kpc) & (d) &  & (mJy)  &  & &  &  &  &  & (d)   \\
\hline
\endhead
\hline
\endfoot
\hline
\endlastfoot
    J0151$-$0635 & 1.465  & 25.7  & 25.0  & 1442  & 5     & 0.70  $\pm$ 3.332  &       &       &       &       &       &       & 0.9   &  \\
    J0304$+$1932 & 1.388  & 15.7  & 0.7   & 2417  & 6     & 10.97  $\pm$ 2.609  & 2.00  & 0.90  & 0.06  & 1.79  & 1.00  & 0.56  & 0.7   &  \\
    J0457$-$6337 & 2.497  & 27.5  & 1.3   & 1     & 2     & 0.17  $\pm$ 0.047  &       &       &       &       &       &       & 1.1   &  \\
    J0610$-$2100 & 0.004  & 60.7  & 3.3   & 1498  & 45    & 0.30  $\pm$ 0.019  & 0.24  & 0.06  & 0.00  & 0.24  & 0.03  & 0.25  & 7.3   & F-R \\
    J0633$-$2015 & 3.253  & 90.7  & 3.7   & 1     & 2     & 0.15  $\pm$ 0.032  &       &       &       &       &       &       & 17.1  &  \\
    J0837$+$0610 & 1.274  & 12.9  & 0.2   & 4175  & 14    & 2.50  $\pm$ 0.575  & 0.54  & 0.23  & 0.03  & 0.49  & 1.14  & 0.61  & 0.8   &  \\
    J0900$-$3144 & 0.011  & 75.7  & 0.4   & 4844  & 100   & 3.53  $\pm$ 0.180  & 0.11  & 0.05  & 0.00  & 0.09  & 0.05  & 0.22  & 118.9  & S \\
    J0922$+$0638 & 0.431  & 27.3  & 1.9   & 3705  & 26    & 2.96  $\pm$ 0.941  & 2.51  & 0.32  & 0.03  & 2.49  & 0.18  & 0.41  & 0.8   &  \\
    J0931$-$1902 & 0.005  & 41.5  & 3.7   & 20    & 3     & 0.42  $\pm$ 0.114  &       &       &       &       &       &       & 12.7  &  \\
    J0953$+$0755 & 0.253  & 3.0   & 0.2   & 3865  & 7     & 74.17  $\pm$ 9.924  & 0.20  & 0.13  & 0.03  & 0.14  & 37.66  & 1.10  & 0.2   &  \\
    J1001$-$5939 & 7.734  & 113.0  & 1.2   & 129   & 7     & 0.24  $\pm$ 0.032  & 0.39  & 0.13  & 0.08  & 0.36  & 0.00  & 0.16  & 15.6  &  \\
    J1002$-$5919 & 0.714  & 347.2  & 3.2   & 1630  & 44    & 0.16  $\pm$ 0.010  & 0.16  & 0.06  & 0.03  & 0.14  & 0.00  & 0.07  & 259.4  & X \\
    J1017$-$5621 & 0.503  & 438.7  & 3.5   & 1     & 1     & 1.77  $\pm$ 0.384  &       &       &       &       &       &       & 445.7  &  \\
    J1017$-$7156 & 0.002  & 94.2  & 1.8   & 102   & 82    & 0.93  $\pm$ 0.058  & 0.25  & 0.06  & 0.00  & 0.25  & 0.01  & 0.19  & 13.0  & F-R \\
    J1038$+$0032 & 0.029  & 26.6  & 5.9   & 93    & 2     & 0.20  $\pm$ 0.072  &       &       &       &       &       &       & 2.1   &  \\
    J1041$-$1942 & 1.386  & 33.8  & 2.5   & 1026  & 8     & 1.27  $\pm$ 0.333  & 0.69  & 0.26  & 0.04  & 0.63  & 0.12  & 0.36  & 3.7   &  \\
    J1046$+$0304 & 0.326  & 25.3  & 5.8   & 1     & 2     & 2.76  $\pm$ 0.706  &       &       &       &       &       &       & 1.9   &  \\
    J1054$-$5946 & 0.228  & 253.4  & 3.5   & 3052  & 24    & 0.33  $\pm$ 0.027  & 0.86  & 0.08  & 0.03  & 0.85  & 0.00  & 0.09  & 114.8  &  \\
    J1056$-$6258 & 0.422  & 320.3  & 2.6   & 1406  & 8     & 27.06  $\pm$ 2.077  & 0.18  & 0.08  & 0.07  & 0.14  & 0.00  & 0.08  & 198.8  &  \\
    J1101$-$6424 & 0.005  & 207.0  & 2.2   & 1619  & 48    & 0.28  $\pm$ 0.017  & 0.18  & 0.06  & 0.00  & 0.17  & 0.00  & 0.11  & 72.7  & S \\
    J1103$-$5403 & 0.003  & 103.9  & 1.7   & 1257  & 57    & 0.31  $\pm$ 0.022  & 0.37  & 0.07  & 0.00  & 0.36  & 0.00  & 0.17  & 15.3  & F-R \\
    J1105$-$4353 & 0.351  & 38.3  & 25.0  & 313   & 2     & 0.19  $\pm$ 0.046  &       &       &       &       &       &       & 8.2   &  \\
    J1105$-$6107 & 0.063  & 271.2  & 2.4   & 4162  & 136   & 1.08  $\pm$ 0.072  & 0.41  & 0.07  & 0.02  & 0.40  & 0.00  & 0.09  & 133.4  & S \\
    J1110$-$5637 & 0.558  & 262.6  & 2.4   & 1     & 1     & 3.47  $\pm$ 0.834  &       &       &       &       &       &       & 126.9  &  \\
    J1114$-$6100 & 0.881  & 677.0  & 5.5   & 2456  & 20    & 2.35  $\pm$ 0.799  & 1.53  & 0.34  & 0.12  & 1.48  & 0.00  & 0.05  & 1383.3  &  \\
    J1119$-$6127 & 0.408  & 704.8  & 6.4   & 4136  & 105   & 1.09  $\pm$ 0.071  & 0.62  & 0.06  & 0.04  & 0.61  & 0.00  & 0.04  & 1626.6  & X \\
    J1125$-$5825 & 0.003  & 124.8  & 1.7   & 3275  & 205   & 1.00  $\pm$ 0.056  & 0.25  & 0.06  & 0.00  & 0.25  & 0.01  & 0.15  & 80.3  & S \\
    J1125$-$6014 & 0.003  & 53.0  & 1.0   & 4843  & 219   & 0.86  $\pm$ 0.061  & 0.79  & 0.07  & 0.00  & 0.79  & 0.02  & 0.27  & 3.0   & F-R \\
    J1136$-$5525 & 0.365  & 85.5  & 1.5   & 1765  & 5     & 4.29  $\pm$ 0.596  &       &       &       &       &       &       & 9.2   &  \\
    J1151$-$6108 & 0.102  & 217.0  & 2.1   & 1680  & 34    & 0.09  $\pm$ 0.006  & 0.16  & 0.06  & 0.01  & 0.15  & 0.00  & 0.10  & 81.2  & F-R \\
    J1210$-$6550 & 4.237  & 37.0  & 0.9   & 1014  & 29    & 0.19  $\pm$ 0.029  & 0.99  & 0.15  & 0.06  & 0.98  & 0.05  & 0.34  & 1.5   &  \\
    J1216$-$6410 & 0.004  & 47.4  & 1.1   & 1560  & 51    & 1.09  $\pm$ 0.059  & 0.20  & 0.05  & 0.00  & 0.20  & 0.03  & 0.29  & 2.6   & F-R \\
    J1224$-$6407 & 0.216  & 97.7  & 1.5   & 3857  & 227   & 7.39  $\pm$ 0.417  & 0.41  & 0.06  & 0.05  & 0.41  & 0.01  & 0.18  & 10.1  & F-R \\
    J1227$-$6208 & 0.035  & 363.0  & 9.5   & 2477   & 97     & 0.26  $\pm$ 0.013  & 0.09  & 0.05  & 0.01  & 0.08  & 0.00  & 0.07  & 491.9  & F-N \\
    J1235$-$54 & 0.638  & 100.0  & 1.7   & 479   & 26    & 0.75  $\pm$ 0.078  & 0.39  & 0.10  & 0.08  & 0.37  & 0.01  & 0.18  & 14.4  &  \\
    J1239$+$2453 & 1.382  & 9.3   & 0.8   & 3260  & 8     & 6.85  $\pm$ 1.546  & 0.38  & 0.23  & 0.10  & 0.29  & 2.67  & 0.70  & 0.1   &  \\
    J1244$-$6359 & 0.147  & 286.5  & 8.5   & 1559  & 35    & 0.13  $\pm$ 0.007  & 0.14  & 0.06  & 0.01  & 0.12  & 0.00  & 0.08  & 284.4  & F-R \\
    J1337$-$6423 & 0.009  & 259.2  & 5.9   & 1421  & 57    & 0.30  $\pm$ 0.017  & 0.16  & 0.06  & 0.00  & 0.14  & 0.00  & 0.09  & 222.4  & F-R\\
    J1405$-$4656 & 0.008  & 13.9  & 3.9   & 824   & 12    & 0.80  $\pm$ 0.068  & 0.30  & 0.09  & 0.00  & 0.29  & 0.68  & 0.59  & 0.3   &  \\
    J1420$-$5625 & 0.034  & 64.6  & 1.3   & 1775  & 47    & 0.18  $\pm$ 0.012  & 0.28  & 0.07  & 0.01  & 0.27  & 0.02  & 0.24  & 5.2   & F-R \\
    J1421$-$4409 & 0.006  & 54.6  & 2.2   & 1210  & 34    & 1.02  $\pm$ 0.093  & 0.61  & 0.09  & 0.00  & 0.60  & 0.04  & 0.27  & 13.6  & F-R \\
    J1431$-$4715 & 0.002  & 59.4  & 1.8   & 2455  & 73    & 0.45  $\pm$ 0.038  & 0.46  & 0.08  & 0.00  & 0.45  & 0.02  & 0.25  & 5.1   & F-R \\
    J1431$-$5740 & 0.004  & 131.2  & 3.6   & 2550  & 90    & 0.36  $\pm$ 0.020  & 0.46  & 0.06  & 0.00  & 0.45  & 0.00  & 0.15  & 36.0  & F-R \\
    J1435$-$6100 & 0.009  & 113.7  & 2.8   & 1814  & 70    & 0.30  $\pm$ 0.020  & 0.30  & 0.07  & 0.00  & 0.30  & 0.00  & 0.16  & 23.8  & S \\
    J1437$-$5959 & 0.062  & 549.6  & 8.5   & 1     & 1     & 0.09  $\pm$ 0.006  &       &       &       &       &       &       & 1115.2  &  \\
    J1439$-$5501 & 0.029  & 14.6  & 0.7   & 4829  & 42    & 0.41  $\pm$ 0.071  & 1.11  & 0.17  & 0.01  & 1.10  & 0.68  & 0.57  & 0.3   & F-DR \\
    J1454$-$5846 & 0.045  & 116.0  & 3.0   & 1814  & 48    & 0.29  $\pm$ 0.016  & 0.15  & 0.06  & 0.01  & 0.14  & 0.00  & 0.16  & 25.6  & F-R \\
    J1502$-$6752 & 0.027  & 151.2  & 2.3   & 1370  & 40    & 0.56  $\pm$ 0.061  & 0.78  & 0.11  & 0.01  & 0.77  & 0.00  & 0.13  & 38.3  & I \\
    J1517$-$4636 & 0.887  & 127.0  & 7.7   & 311   & 11    & 0.29  $\pm$ 0.023  & 0.38  & 0.08  & 0.09  & 0.36  & 0.01  & 0.15  & 49.5  &  \\
    J1519$-$5734 & 0.519  & 664.0  & 6.9   & 1     & 2     & 0.59  $\pm$ 0.081  &       &       &       &       &       &       & 1485.1  &  \\
    J1530$-$6343 & 0.910  & 201.2  & 10.8  & 350   & 7     & 0.36  $\pm$ 0.036  & 0.21  & 0.10  & 0.07  & 0.17  & 0.00  & 0.11  & 153.0  &  \\
    J1532$-$56 & 0.523  & 282.0  & 2.3   & 1489  & 24    & 0.09  $\pm$ 0.005  & 0.13  & 0.06  & 0.02  & 0.11  & 0.00  & 0.09  & 198.5  &  \\
    J1537$-$5312 & 0.007  & 117.5  & 3.1   & 787   & 19    & 0.20  $\pm$ 0.015  & 0.19  & 0.07  & 0.00  & 0.17  & 0.00  & 0.16  & 26.6  &  \\
    J1538$-$5732 & 0.341  & 152.7  & 3.4   & 1     & 2     & 0.21  $\pm$ 0.038  &       &       &       &       &       &       & 48.5  &  \\
    J1543$-$0620 & 0.709  & 18.4  & 1.1   & 4030  & 14    & 2.30  $\pm$ 0.484  & 1.01  & 0.21  & 0.10  & 0.98  & 1.45  & 0.51  & 0.6   &  \\
    J1543$-$5149 & 0.002  & 50.9  & 6.0   & 2018  & 34    & 0.60  $\pm$ 0.047  & 0.22  & 0.08  & 0.00  & 0.20  & 0.09  & 0.28  & 28.4  & F-R \\
    J1545$-$4550 & 0.004  & 68.4  & 2.2   & 873   & 45    & 0.79  $\pm$ 0.049  & 0.23  & 0.06  & 0.00  & 0.22  & 0.01  & 0.23  & 7.6   & S \\
    J1546$-$59 & 0.008  & 168.3  & 3.9   & 422   & 11    & 0.18  $\pm$ 0.011  & 0.11  & 0.06  & 0.00  & 0.09  & 0.00  & 0.12  & 63.2  &  \\
    J1551$-$6214 & 0.199  & 122.2  & 1.4   & 404   & 15    & 0.33  $\pm$ 0.032  & 0.25  & 0.10  & 0.04  & 0.23  & 0.01  & 0.15  & 32.6  &  \\
    J1612$-$55 & 0.847  & 312.0  & 6.9   & 1249  & 28    & 0.10  $\pm$ 0.007  & 0.15  & 0.07  & 0.03  & 0.14  & 0.00  & 0.08  & 306.3  &  \\
    J1614$-$3846 & 0.464  & 111.0  & 5.6   & 262   & 5     & 0.21  $\pm$ 0.020  &       &       &       &       &       &       & 32.1  &  \\
    J1616$-$5017 & 0.491  & 194.0  & 4.4   & 672   & 24    & 0.20  $\pm$ 0.012  & 0.14  & 0.06  & 0.04  & 0.12  & 0.00  & 0.11  & 80.4  &  \\
    J1622$-$4950 & 4.326  & 881.0  & 5.7   & 454   & 103   & 15.97  $\pm$ 1.104  & 0.35  & 0.07  & 0.16  & 0.31  & 0.00  & 0.04  & 2456.6  & I \\
    J1622$-$6617 & 0.024  & 88.0  & 4.0   & 1330   & 67    & 0.43  $\pm$ 0.040  & 0.64  & 0.09  & 0.01  & 0.63  & 0.01  & 0.19  & 39.4  & F-N \\
    J1627$-$49 & 0.624  & 594.0  & 5.5   & 721   & 20    & 0.14  $\pm$ 0.009  & 0.10  & 0.06  & 0.02  & 0.08  & 0.00  & 0.05  & 1057.3  &  \\
    J1627$-$51 & 0.440  & 201.0  & 3.7   & 748   & 22    & 0.17  $\pm$ 0.011  & 0.12  & 0.06  & 0.03  & 0.10  & 0.00  & 0.11  & 89.9  &  \\
    J1634$-$5107 & 0.507  & 372.8  & 12.5  & 3652  & 28    & 0.75  $\pm$ 0.067  & 0.30  & 0.09  & 0.06  & 0.28  & 0.00  & 0.07  & 453.9  &  \\
    J1637$-$4450 & 0.253  & 470.7  & 11.4  & 2     & 2     & 0.37  $\pm$ 0.052  &       &       &       &       &       &       & 931.6  &  \\
    J1638$-$4233 & 0.511  & 406.0  & 0.3   & 837   & 5     & 0.24  $\pm$ 0.032  &       &       &       &       &       &       & 952.2  &  \\
    J1638$-$44 & 0.568  & 494.0  & 7.2   & 1177  & 28    & 0.18  $\pm$ 0.010  & 0.11  & 0.06  & 0.03  & 0.09  & 0.00  & 0.06  & 1077.1  &  \\
    J1645$-$0317 & 0.388  & 35.8  & 1.3   & 3395  & 7     & 13.06  $\pm$ 9.511  & 1.77  & 0.73  & 0.02  & 1.61  & 0.05  & 0.35  & 1.3   &  \\
    J1649$-$3805 & 0.262  & 213.8  & 8.8   & 3488  & 4     & 0.85  $\pm$ 0.114  &       &       &       &       &       &       & 156.9  &  \\
    J1651$-$5255 & 0.891  & 164.0  & 5.8   & 1     & 2     & 2.91  $\pm$ 0.503  &       &       &       &       &       &       & 73.2  &  \\
    J1653$-$4030 & 1.019  & 425.0  & 17.2  & 1     & 2     & 0.39  $\pm$ 0.069  &       &       &       &       &       &       & 922.4  &  \\
    J1653$-$45 & 0.951  & 207.0  & 3.5   & 745   & 19    & 0.14  $\pm$ 0.017  & 0.30  & 0.12  & 0.03  & 0.27  & 0.00  & 0.11  & 92.4  &  \\
    J1658$-$47 & 0.369  & 533.0  & 23.4  & 285   & 8     & 0.18  $\pm$ 0.013  & 0.10  & 0.07  & 0.02  & 0.07  & 0.00  & 0.05  & 1731.6  &  \\
    J1701$-$4533 & 0.323  & 526.0  & 19.6  & 32    & 2     & 2.65  $\pm$ 0.533  &       &       &       &       &       &       & 1540.2  &  \\
    J1704$-$5236 & 0.231  & 170.5  & 0.7   & 405   & 12    & 0.51  $\pm$ 0.034  & 0.13  & 0.07  & 0.04  & 0.11  & 0.00  & 0.12  & 101.6  &  \\
    J1708$-$3506 & 0.005  & 146.7  & 3.3   & 3076  & 172   & 1.18  $\pm$ 0.060  & 0.13  & 0.05  & 0.00  & 0.12  & 0.00  & 0.14  & 147.7  & I \\
    J1709$-$4401 & 0.865  & 220.6  & 4.5   & 2400  & 37    & 1.09  $\pm$ 0.088  & 0.32  & 0.08  & 0.04  & 0.31  & 0.00  & 0.10  & 120.0  & I \\
    J1711$-$4322 & 0.103  & 191.5  & 4.0   & 1538  & 43    & 0.33  $\pm$ 0.017  & 0.09  & 0.05  & 0.01  & 0.07  & 0.00  & 0.11  & 83.7  & F-N \\
    J1715$-$3859 & 0.928  & 817.0  & 5.2   & 1     & 2     & 0.56  $\pm$ 0.072  &       &       &       &       &       &       & 1998.7  &  \\
    J1716$-$4005 & 0.312  & 435.0  & 5.4   & 1701  & 2     & 0.81  $\pm$ 0.116  &       &       &       &       &       &       & 543.8  &  \\
    J1718$-$3718 & 3.379  & 371.1  & 9.5   & 3520  & 12    & 0.14  $\pm$ 0.011  & 0.29  & 0.08  & 0.10  & 0.26  & 0.00  & 0.07  & 331.4  &  \\
    J1718$-$41 & 0.548  & 354.0  & 3.9   & 685   & 15    & 0.11  $\pm$ 0.008  & 0.14  & 0.07  & 0.02  & 0.12  & 0.00  & 0.07  & 516.5  &  \\
    J1719$-$1438 & 0.006  & 36.9  & 0.3   & 2806  & 84    & 0.34  $\pm$ 0.028  & 0.53  & 0.08  & 0.00  & 0.53  & 0.04  & 0.35  & 0.9   & F-R \\
    J1727$-$2946 & 0.027  & 60.9  & 1.9   & 1321  & 35    & 0.22  $\pm$ 0.015  & 0.32  & 0.06  & 0.01  & 0.31  & 0.02  & 0.25  & 5.5   & F-R \\
    J1729$-$2117 & 0.066  & 35.0  & 1.0   & 2174  & 28    & 0.11  $\pm$ 0.021  & 0.84  & 0.18  & 0.01  & 0.81  & 0.07  & 0.36  & 1.4   &  \\
    J1730$-$34 & 0.100  & 628.0  & 4.6   & 1     & 1     & 0.23  $\pm$ 0.005  &       &       &       &       &       &       & 1078.5  &  \\
    J1731$-$1847 & 0.002  & 106.5  & 4.8   & 2904  & 123   & 0.39  $\pm$ 0.027  & 0.46  & 0.07  & 0.00  & 0.45  & 0.01  & 0.17  & 27.1  & F-R \\
    J1732$-$35 & 0.127  & 340.0  & 4.0   & 603   & 10    & 0.13  $\pm$ 0.009  & 0.20  & 0.07  & 0.02  & 0.19  & 0.00  & 0.07  & 280.2  &  \\
    J1735$-$0724 & 0.419  & 73.5  & 0.2   & 1     & 2     & 3.19  $\pm$ 0.648  &       &       &       &       &       &       & 1.2   &  \\
    J1740$+$1000 & 0.154  & 23.9  & 1.2   & 3394  & 9     & 2.43  $\pm$ 1.308  & 1.24  & 0.54  & 0.01  & 1.12  & 0.17  & 0.45  & 0.8   &  \\
    J1741$-$34 & 0.875  & 241.0  & 4.6   & 818   & 14    & 0.10  $\pm$ 0.007  & 0.33  & 0.07  & 0.02  & 0.32  & 0.00  & 0.10  & 144.7  &  \\
    J1743$-$35 & 0.570  & 174.0  & 11.6  & 685   & 15    & 0.11  $\pm$ 0.007  & 0.16  & 0.07  & 0.03  & 0.14  & 0.00  & 0.12  & 68.8  &  \\
    J1744$-$5337 & 0.356  & 124.4  & 19.3  & 626   & 37    & 0.29  $\pm$ 0.020  & 0.23  & 0.07  & 0.05  & 0.21  & 0.01  & 0.15  & 75.1  & I \\
    J1745$-$2758 & 0.488  & 422.0  & 4.2   & 1     & 1     & 0.17  $\pm$ 0.015  &       &       &       &       &       &       & 446.7  &  \\
    J1747$-$4036 & 0.002  & 153.0  & 7.2   & 1798  & 14    & 1.34  $\pm$ 0.077  & 0.11  & 0.06  & 0.00  & 0.10  & 0.00  & 0.13  & 260.6  &  \\
    J1748$-$30 & 0.383  & 584.0  & 13.8  & 1     & 1     & 0.16  $\pm$ 0.016  &       &       &       &       &       &       & 1611.3  &  \\
    J1750$-$28 & 1.301  & 388.0  & 4.1   & 1     & 1     & 0.15  $\pm$ 0.012  &       &       &       &       &       &       & 370.8  &  \\
    J1753$-$2240 & 0.095  & 158.6  & 3.2   & 1475  & 40    & 0.14  $\pm$ 0.010  & 0.26  & 0.07  & 0.01  & 0.25  & 0.00  & 0.13  & 50.9  & S \\
    J1754$-$3510 & 0.393  & 82.3  & 2.6   & 1     & 2     & 1.04  $\pm$ 0.207  &       &       &       &       &       &       & 11.8  &  \\
    J1755$-$26 & 0.431  & 405.0  & 3.9   & 164   & 2     & 0.16  $\pm$ 0.023  &       &       &       &       &       &       & 399.5  &  \\
    J1756$-$25 & 0.856  & 706.0  & 4.8   & 1     & 1     & 0.22  $\pm$ 0.023  &       &       &       &       &       &       & 1409.2  &  \\
    J1757$-$2223 & 0.185  & 239.3  & 3.7   & 4     & 3     & 0.46  $\pm$ 0.072  &       &       &       &       &       &       & 129.0  &  \\
    J1759$-$24 & 1.458  & 772.0  & 4.9   & 1248  & 4     & 0.28  $\pm$ 0.055  &       &       &       &       &       &       & 1718.0  &  \\
    J1801$-$1417 & 0.004  & 57.3  & 1.1   & 1272  & 26    & 1.36  $\pm$ 0.135  & 0.49  & 0.10  & 0.00  & 0.48  & 0.05  & 0.26  & 18.9  &  \\
    J1801$-$3210 & 0.007  & 177.7  & 6.1   & 3201  & 113   & 0.35  $\pm$ 0.018  & 0.17  & 0.05  & 0.00  & 0.16  & 0.00  & 0.12  & 67.6  & F-R \\
    J1802$-$2124 & 0.013  & 149.6  & 3.0   & 4282  & 45    & 0.95  $\pm$ 0.054  & 0.16  & 0.06  & 0.00  & 0.15  & 0.01  & 0.13  & 745.8  & S \\
    J1804$-$2858 & 0.001  & 232.0  & 8.4   & 1755  & 63    & 0.29  $\pm$ 0.022  & 0.28  & 0.08  & 0.00  & 0.27  & 0.00  & 0.10  & 181.1  & S \\
    J1807$-$2715 & 0.828  & 313.0  & 15.4  & 3808  & 30    & 1.11  $\pm$ 0.060  & 0.15  & 0.05  & 0.11  & 0.08  & 0.00  & 0.08  & 459.3  & F-J \\
    J1809$-$1943 & 5.541  & 178.0  & 3.2   & 842   & 33    & 0.66  $\pm$ 1.229  & 6.01  & 1.85  & 0.07  & 5.72  & 0.00  & 0.12  & 83.7  & X \\
    J1811$-$2405 & 0.003  & 60.6  & 1.8   & 1605  & 59    & 1.37  $\pm$ 0.078  & 0.22  & 0.06  & 0.00  & 0.21  & 0.02  & 0.25  & 5.4   & F-R \\
    J1811$-$4930 & 1.433  & 44.0  & 1.4   & 316   & 3     & 0.37  $\pm$ 0.072  &       &       &       &       &       &       & 2.6   &  \\
    J1818$-$1422 & 0.291  & 622.0  & 5.5   & 3808  & 23    & 4.24  $\pm$ 0.214  & 0.07  & 0.05  & 0.06  & nan & 0.00  & 0.05  & 1155.4  &  \\
    J1819$-$17 & 2.352  & 67.0  & 1.9   & 1     & 1     & 0.22 $\pm$ 0.014  &       &       &       &       &       &       & 6.6   &  \\
    J1820$-$0509 & 0.337  & 104.0  & 3.3   & 1     & 1     & 0.71  $\pm$ 0.010  &       &       &       &       &       &       & 21.6  &  \\
    J1824$-$0127 & 2.499  & 58.0  & 1.9   & 1     & 2     & 0.37  $\pm$ 0.090  &       &       &       &       &       &       & 5.0   &  \\
    J1824$-$1350 & 1.397  & 551.0  & 4.7   & 5     & 2     & 0.14  $\pm$ 0.019  &       &       &       &       &       &       & 831.0  &  \\
    J1828$-$2119 & 0.515  & 268.0  & 18.0  & 1     & 2     & 0.61  $\pm$ 0.095  &       &       &       &       &       &       & 358.8  &  \\
    J1830$-$0131 & 0.153  & 95.7  & 3.5   & 1     & 2     & 0.48  $\pm$ 0.092  &       &       &       &       &       &       & 18.7  &  \\
    J1831$-$0952 & 0.067  & 247.0  & 3.7   & 187   & 11    & 0.32  $\pm$ 0.028  & 0.17  & 0.09  & 0.01  & 0.15  & 0.00  & 0.09  & 137.0  &  \\
    J1834$-$09 & 0.512  & 529.0  & 4.9   & 1     & 1     & 0.28  $\pm$ 0.033  &       &       &       &       &       &       & 775.6  &  \\
    J1838$-$01 & 0.183  & 320.4  & 7.8   & 1     & 1     & 0.12  $\pm$0.016  &       &       &       &       &       &       & 343.2  &  \\
    J1843$-$0702 & 0.192  & 228.4  & 4.3   & 3792  & 31    & 0.25  $\pm$ 0.016  & 0.18  & 0.06  & 0.03  & 0.16  & 0.00  & 0.10  & 125.4  & X \\
    J1845$-$1114 & 0.206  & 206.7  & 6.7   & 1     & 2     & 0.51  $\pm$0.084  &       &       &       &       &       &       & 127.5  &  \\
    J1846$-$4249 & 2.273  & 62.0  & 3.7   & 208   & 4     & 0.38  $\pm$ 0.059  &       &       &       &       &       &       & 8.0   &  \\
    J1850$-$0026 & 0.167  & 947.0  & 6.7   & 1     & 1     & 0.55  $\pm$ 0.016  &       &       &       &       &       &       & 3096.9  &  \\
    J1856$-$0526 & 0.370  & 130.5  & 4.0   & 1     & 2     & 0.83  $\pm$ 0.151  &       &       &       &       &       &       & 37.8  &  \\
    J1901$+$0621 & 0.832  & 94.0  & 2.9   & 1     & 2     & 0.30  $\pm$ 0.060  &       &       &       &       &       &       & 16.5  &  \\
    J1902$+$0615 & 0.674  & 502.9  & 7.2   & 4097  & 13    & 1.77  $\pm$ 0.124  & 0.19  & 0.07  & 0.06  & 0.17  & 0.00  & 0.06  & 951.1  &  \\
    J1903$+$0925 & 0.357  & 162.0  & 6.3   & 30    & 2     & 0.85  $\pm$ 0.155  &       &       &       &       &       &       & 74.0  &  \\
    J1903$-$7051 & 0.004  & 19.7  & 0.9   & 151   & 8     & 0.58  $\pm$ 0.144  & 0.56  & 0.25  & 0.00  & 0.50  & 0.59  & 0.50  & 1.9   &  \\
    J1910$-$5959A & 0.003  & 33.7  & 1.7   & 1793  & 41    & 0.22  $\pm$ 0.017  & 0.72  & 0.08  & 0.00  & 0.72  & 0.08  & 0.37  & 4.7   & S \\
    J1910$-$5959B & 0.008  & 33.3  & 1.6   & 1     & 1     & 0.11  $\pm$ 0.010  &       &       &       &       &       &       & 1.6   &  \\
    J1910$-$5959C & 0.005  & 33.2  & 1.6   & 147   & 4     & 0.26  $\pm$ 0.038  &       &       &       &       &       &       & 10.3  &  \\
    J1911$+$1347 & 0.005  & 31.0  & 1.4   & 1030  & 18    & 0.45  $\pm$ 0.037  & 0.16  & 0.08  & 0.00  & 0.14  & 0.48  & 0.38  & 13.4  &  \\
    J1921$+$0812 & 0.211  & 84.0  & 2.9   & 1     & 2     & 0.62  $\pm$ 0.123  &       &       &       &       &       &       & 13.0  &  \\
    J1932$+$1059 & 0.227  & 3.2   & 0.2   & 1358  & 6     & 29.93  $\pm$ 67.763  & 4.26  & 2.26  & 0.04  & 3.61  & 23.63  & 1.07  & 0.1   &  \\
    J1932$+$2020 & 0.268  & 211.2  & 5.0   & 3751  & 8     & 1.20  $\pm$ 0.070  & 0.10  & 0.06  & 0.04  & 0.08  & 0.00  & 0.10  & 114.6  &  \\
    J1933$-$6211 & 0.004  & 11.5  & 0.7   & 2513  & 51    & 0.86  $\pm$ 0.187  & 2.47  & 0.22  & 0.00  & 2.46  & 1.93  & 0.64  & 1.6   & F-D \\
    J1935$+$1616 & 0.359  & 158.5  & 4.3   & 1     & 1     & 30.05  $\pm$ 0.498  &       &       &       &       &       &       & 62.3  &  \\
    J1938$+$2213 & 0.166  & 91.0  & 3.4   & 969   & 4     & 0.62  $\pm$ 0.108  &       &       &       &       &       &       & 16.6  &  \\
    J1944$+$0907 & 0.005  & 24.4  & 1.2   & 1     & 2     & 2.69  $\pm$ 0.749  &       &       &       &       &       &       & 1.6   &  \\
    J2010$-$1323 & 0.005  & 22.2  & 1.2   & 1     & 2     & 0.66  $\pm$ 0.173  &       &       &       &       &       &       & 3.7   &  \\
    J2053$-$7200 & 0.341  & 17.3  & 4.0   & 2862  & 15    & 2.42  $\pm$ 1.051  & 1.06  & 0.43  & 0.04  & 0.96  & 0.83  & 0.53  & 0.4   &  \\
    J2108$-$3429 & 1.423  & 30.2  & 3.9   & 1     & 2     & 0.28  $\pm$ 0.071  &       &       &       &       &       &       & 2.1   &  \\
    J2144$-$3933 & 8.510  & 3.4   & 22.2  & 3703  & 5     & 0.61  $\pm$ 0.172  &       &       &       &       &       &       & 0.1   &  \\
    J2234$+$0944 & 0.004  & 17.8  & 1.6   & 1566  & 62    & 1.10  $\pm$ 0.136  & 1.36  & 0.12  & 0.00  & 1.35  & 0.67  & 0.52  & 1.5   & F-DR \\
    J2236$-$5526 & 0.007  & 20.2  & 2.1   & 1237  & 51    & 0.25  $\pm$ 0.036  & 1.28  & 0.14  & 0.00  & 1.28  & 0.33  & 0.49  & 0.8   & F-DR \\
    J2322$-$2650 & 0.003  & 6.2   & 0.8   & 1868  & 43    & 0.15  $\pm$ 0.012  & 0.50  & 0.08  & 0.00  & 0.49  & 2.41  & 0.83  & 0.1   & F-D \\
\end{longtable}
}


\bsp	
\label{lastpage}
\end{document}